\pdfoutput=1
\documentclass[12pt,a4paper]{article}

\usepackage{color}

\usepackage{ifthen} 
\newboolean{pdflatex}
\setboolean{pdflatex}{true} 
\newboolean{articletitles}
\setboolean{articletitles}{true} 
\newboolean{uprightparticles}
\setboolean{uprightparticles}{true} 
\newboolean{inbibliography}
\setboolean{inbibliography}{false} 

\newboolean{prl}
\setboolean{prl}{False} 

\usepackage{lscape}
\usepackage{graphpap} 
\usepackage{multirow} 
\usepackage{rotating} 
\usepackage{mathrsfs}
\usepackage{mathtools} 
\usepackage{dashrule}
\usepackage{tikz}
\usepackage{enumerate}
\usetikzlibrary{patterns}

\def\paperauthors{LHCb collaboration} 

\newcommand{\lambdabpipi}{{\ensuremath{\Lb \pip \pim}}\xspace}
\newcommand{\lambdabpipiSS}{{\ensuremath{\Lb \Ppi^\pm \Ppi^\pm}}\xspace}

\newcommand{\lblcpi}{\ensuremath{\decay{\Lb}{\Lc\pim}}\xspace}

\newcommand{\lbjpsipk}{\ensuremath{\decay{\Lb}{\jpsi\proton\Km}}\xspace}

\newcommand{\lboneplow}{\ensuremath{\Lambdares_{\bquark}\mathrm{(5912)}^0}\xspace}
\newcommand{\lbonephigh}{\ensuremath{\Lambdares_{\bquark}\mathrm{(5920)}^0}\xspace}
\newcommand{\lbonedlow}{\ensuremath{\Lambdares_{\bquark}\mathrm{(6146)}^0}\xspace}
\newcommand{\lbonedhigh}{\ensuremath{\Lambdares_{\bquark}\mathrm{(6152)}^0}\xspace}
\newcommand{\lbstarstar}{\ensuremath{\Lambdares_{\bquark}^{\ast\ast 0}}\xspace}

\newcommand{\Sigmab}{\ensuremath{\Sigmares_{\bquark}}\xspace}
\newcommand{\Sigmabstar}{\ensuremath{\Sigmares^{\ast}_{\bquark}}\xspace}

\def\paperauthors{LHCb collaboration} 

\def\paperasciititle{Observation of a new baryon state in the Lbpi+pi- mass spectrum} 
\def\papertitle{Observation of a~new baryon state in~the~$\Lb\pip\pim$~mass spectrum} 
\def\paperkeywords{{High Energy Physics}, {LHCb}} 
\def\papercopyright{\the\year\ CERN for the benefit of the LHCb collaboration} 
\def\paperlicence{CC BY 4.0 licence}
\def\paperlicenceurl{https://creativecommons.org/licenses/by/4.0/}

\makeatletter
\g@addto@macro\bfseries{\boldmath}
\makeatother

\usepackage[top=1in, bottom=1.25in, left=1in, right=1in]{geometry}

\usepackage{microtype}
\usepackage{lineno}  
\usepackage{xspace} 
\usepackage{caption} 

\usepackage{graphicx}  
\usepackage{color}
\usepackage{colortbl}
\graphicspath{{./figs/}} 

\usepackage{amsmath} 
\usepackage{amssymb}
\usepackage{amsfonts}
\usepackage{upgreek} 

\newcommand*\patchAmsMathEnvironmentForLineno[1]{%
\expandafter\let\csname old#1\expandafter\endcsname\csname #1\endcsname
\expandafter\let\csname oldend#1\expandafter\endcsname\csname
end#1\endcsname
 \renewenvironment{#1}%
   {\linenomath\csname old#1\endcsname}%
   {\csname oldend#1\endcsname\endlinenomath}%
}
\newcommand*\patchBothAmsMathEnvironmentsForLineno[1]{%
  \patchAmsMathEnvironmentForLineno{#1}%
  \patchAmsMathEnvironmentForLineno{#1*}%
}
\AtBeginDocument{%
\patchBothAmsMathEnvironmentsForLineno{equation}%
\patchBothAmsMathEnvironmentsForLineno{align}%
\patchBothAmsMathEnvironmentsForLineno{flalign}%
\patchBothAmsMathEnvironmentsForLineno{alignat}%
\patchBothAmsMathEnvironmentsForLineno{gather}%
\patchBothAmsMathEnvironmentsForLineno{multline}%
\patchBothAmsMathEnvironmentsForLineno{eqnarray}%
}

\usepackage{hyperxmp}

\usepackage[pdftex,
            pdfauthor={\paperauthors},
            pdftitle={\paperasciititle},
            pdfkeywords={\paperkeywords},
            pdfcopyright={Copyright (C) \papercopyright},
            pdflicenseurl={\paperlicenceurl}]{hyperref}

\usepackage[colorinlistoftodos,textsize=scriptsize]{todonotes}

\usepackage[bottom,flushmargin,hang,multiple]{footmisc}

\usepackage[all]{hypcap} 

\usepackage{xspace} 
\usepackage{upgreek}

\def\lhcb   {\mbox{LHCb}\xspace}

\def\cdf    {\mbox{CDF}\xspace}

\def\MagUp {\mbox{\em Mag\kern -0.05em Up}\xspace}

\ifthenelse{\boolean{uprightparticles}}%
{

 \def\Pmu         {\ensuremath{\upmu}\xspace}

 \def\Ppi         {\ensuremath{\uppi}\xspace}                 
                  
 \def\Prho        {\ensuremath{\uprho}\xspace}

 \def\Pchi        {\ensuremath{\upchi}\xspace}                 
 \def\Ppsi        {\ensuremath{\uppsi}\xspace}

 \def\PDelta      {\ensuremath{\Delta}\xspace}                 
 \def\PXi      {\ensuremath{\Xi}\xspace}                 
 \def\PLambda      {\ensuremath{\Lambda}\xspace}                 
 \def\PSigma      {\ensuremath{\Sigma}\xspace}                 
 \def\POmega      {\ensuremath{\Omega}\xspace}                 
 \def\PUpsilon      {\ensuremath{\Upsilon}\xspace}                 
 
 \def\PB      {\ensuremath{\mathrm{B}}\xspace}                 
                  
 \def\PD      {\ensuremath{\mathrm{D}}\xspace}

 \def\PJ      {\ensuremath{\mathrm{J}}\xspace}                 
 \def\PK      {\ensuremath{\mathrm{K}}\xspace}

 \def\Pb      {\ensuremath{\mathrm{b}}\xspace}                 
 \def\Pc      {\ensuremath{\mathrm{c}}\xspace}                 
 \def\Pd      {\ensuremath{\mathrm{d}}\xspace}

 \def\Pi      {\ensuremath{\mathrm{i}}\xspace}

 \def\Pp      {\ensuremath{\mathrm{p}}\xspace}                 
 \def\Pq      {\ensuremath{\mathrm{q}}\xspace}

 \def\Pu      {\ensuremath{\mathrm{u}}\xspace}

}
{

 \def\Pmu         {\ensuremath{\mu}\xspace}

 \def\Ppi         {\ensuremath{\pi}\xspace}                 
                  
 \def\Prho        {\ensuremath{\rho}\xspace}

 \def\Pchi        {\ensuremath{\chi}\xspace}                 
 \def\Ppsi        {\ensuremath{\psi}\xspace}                 
                  
 \mathchardef\PDelta="7101
 \mathchardef\PXi="7104
 \mathchardef\PLambda="7103
 \mathchardef\PSigma="7106
 \mathchardef\POmega="710A
 \mathchardef\PUpsilon="7107
                  
 \def\PB      {\ensuremath{B}\xspace}                 
                  
 \def\PD      {\ensuremath{D}\xspace}

 \def\PJ      {\ensuremath{J}\xspace}                 
 \def\PK      {\ensuremath{K}\xspace}

 \def\Pb      {\ensuremath{b}\xspace}                 
 \def\Pc      {\ensuremath{c}\xspace}                 
 \def\Pd      {\ensuremath{d}\xspace}

 \def\Pi      {\ensuremath{i}\xspace}

 \def\Pp      {\ensuremath{p}\xspace}                 
 \def\Pq      {\ensuremath{q}\xspace}

 \def\Pu      {\ensuremath{u}\xspace}

}

\makeatletter
\ifcase \@ptsize \relax
  \newcommand{\miniscule}{\@setfontsize\miniscule{4}{5}}
\or
  \newcommand{\miniscule}{\@setfontsize\miniscule{5}{6}}
\or
  \newcommand{\miniscule}{\@setfontsize\miniscule{5}{6}}
\fi
\makeatother

\DeclareRobustCommand{\optbar}[1]{\shortstack{{\miniscule (\rule[.5ex]{1.25em}{.18mm})}
  \\ [-.7ex] $#1$}}

\def\mumu       {{\ensuremath{\Pmu^+\Pmu^-}}\xspace}

\def\quark     {{\ensuremath{\Pq}}\xspace}

\def\uquark    {{\ensuremath{\Pu}}\xspace}

\def\dquark    {{\ensuremath{\Pd}}\xspace}

\def\cquark    {{\ensuremath{\Pc}}\xspace}

\def\bquark    {{\ensuremath{\Pb}}\xspace}

\def\pion   {{\ensuremath{\Ppi}}\xspace}

\def\pip    {{\ensuremath{\pion^+}}\xspace}
\def\pim    {{\ensuremath{\pion^-}}\xspace}
\def\pipm   {{\ensuremath{\pion^\pm}}\xspace}

\def\kaon    {{\ensuremath{\PK}}\xspace}
  \def\Kbar    {{\kern 0.2em\overline{\kern -0.2em \PK}{}}\xspace}

\def\KorKbar    {\kern 0.18em\optbar{\kern -0.18em K}{}\xspace}

\def\Kp      {{\ensuremath{\kaon^+}}\xspace}
\def\Km      {{\ensuremath{\kaon^-}}\xspace}

\def\KS      {{\ensuremath{\kaon^0_{\mathrm{ \scriptscriptstyle S}}}}\xspace}

  \def\Dbar    {{\kern 0.2em\overline{\kern -0.2em \PD}{}}\xspace}

\def\DorDbar    {\kern 0.18em\optbar{\kern -0.18em D}{}\xspace}

\def\B       {{\ensuremath{\PB}}\xspace}
\def\Bbar    {{\ensuremath{\kern 0.18em\overline{\kern -0.18em \PB}{}}}\xspace}

\def\BorBbar    {\kern 0.18em\optbar{\kern -0.18em B}{}\xspace}

\def\Bu      {{\ensuremath{\B^+}}\xspace}

\def\jpsi     {{\ensuremath{{\PJ\mskip -3mu/\mskip -2mu\Ppsi\mskip 2mu}}}\xspace}

  \def\Y#1S{\ensuremath{\PUpsilon{(#1S)}}\xspace}

\def\proton      {{\ensuremath{\Pp}}\xspace}

\def\Lz          {{\ensuremath{\PLambda}}\xspace}
\def\Lbar        {{\ensuremath{\kern 0.1em\overline{\kern -0.1em\PLambda}}}\xspace}
\def\LorLbar    {\kern 0.18em\optbar{\kern -0.18em \PLambda}{}\xspace}
\def\Lambdares   {{\ensuremath{\PLambda}}\xspace}

\def\Sigmares    {{\ensuremath{\PSigma}}\xspace}

\def\Lb      {{\ensuremath{\Lz^0_\bquark}}\xspace}

\def\Lc      {{\ensuremath{\Lz^+_\cquark}}\xspace}

\newcommand{\decay}[2]{\mbox{\ensuremath{#1\!\to #2}}\xspace}         

\def\to                 {\ensuremath{\rightarrow}\xspace}

\def\AT#1     {\ensuremath{A_{\mathrm{T}}^{#1}}\xspace}           

\def\C#1      {\ensuremath{\mathcal{C}_{#1}}\xspace}                       
\def\Cp#1     {\ensuremath{\mathcal{C}_{#1}^{'}}\xspace}                    
\def\Ceff#1   {\ensuremath{\mathcal{C}_{#1}^{\mathrm{(eff)}}}\xspace}        
\def\Cpeff#1  {\ensuremath{\mathcal{C}_{#1}^{'\mathrm{(eff)}}}\xspace}       
\def\Ope#1    {\ensuremath{\mathcal{O}_{#1}}\xspace}                       
\def\Opep#1   {\ensuremath{\mathcal{O}_{#1}^{'}}\xspace}                    

\newcommand{\tev}{\ifthenelse{\boolean{inbibliography}}{\ensuremath{~T\kern -0.05em eV}}{\ensuremath{\mathrm{\,Te\kern -0.1em V}}}\xspace}
\newcommand{\Tev}{\ifthenelse{\boolean{inbibliography}}{\ensuremath{~T\kern -0.05em eV}}{\ensuremath{\mathrm{\,Te\kern -0.1em V}}}\xspace}
\newcommand{\gev}{\ensuremath{\mathrm{\,Ge\kern -0.1em V}}\xspace}
\newcommand{\mev}{\ensuremath{\mathrm{\,Me\kern -0.1em V}}\xspace}
\newcommand{\kev}{\ensuremath{\mathrm{\,ke\kern -0.1em V}}\xspace}
\newcommand{\ev}{\ensuremath{\mathrm{\,e\kern -0.1em V}}\xspace}
\newcommand{\gevc}{\ensuremath{{\mathrm{\,Ge\kern -0.1em V\!/}c}}\xspace}
\newcommand{\mevc}{\ensuremath{{\mathrm{\,Me\kern -0.1em V\!/}c}}\xspace}
\newcommand{\gevcc}{\ensuremath{{\mathrm{\,Ge\kern -0.1em V\!/}c^2}}\xspace}
\newcommand{\gevgevcccc}{\ensuremath{{\mathrm{\,Ge\kern -0.1em V^2\!/}c^4}}\xspace}
\newcommand{\mevcc}{\ensuremath{{\mathrm{\,Me\kern -0.1em V\!/}c^2}}\xspace}
\newcommand{\kevcc}{\ensuremath{{\mathrm{\,ke\kern -0.1em V\!/}c^2}}\xspace}

\def\invfb   {\ensuremath{\mbox{\,fb}^{-1}}\xspace}

\newcommand{\chisq}{\ensuremath{\chi^2}\xspace}

\def\gsim{{~\raise.15em\hbox{$>$}\kern-.85em
          \lower.35em\hbox{$\sim$}~}\xspace}
\def\lsim{{~\raise.15em\hbox{$<$}\kern-.85em
          \lower.35em\hbox{$\sim$}~}\xspace}

\def\sPlot{\mbox{\em sPlot}\xspace}

\def\pt         {\ensuremath{p_{\mathrm{ T}}}\xspace}
\def\ptot       {\ensuremath{p}\xspace}

\def\evtgen     {\mbox{\textsc{EvtGen}}\xspace}

\def\geant      {\mbox{\textsc{Geant4}}\xspace}

\def\photos     {\mbox{\textsc{Photos}}\xspace}

\def\pythia     {\mbox{\textsc{Pythia}}\xspace}

\def\tell1  {TELL1\xspace}
\def\ukl1   {UKL1\xspace}

\newcommand{\eg}{\mbox{\itshape e.g.}\xspace}

\usepackage{cite} 
\usepackage{mciteplus}

\usepackage{rotating}

\begin{document}

\renewcommand{\thefootnote}{\fnsymbol{footnote}}
\setcounter{footnote}{1}


\begin{titlepage}
\pagenumbering{roman}

\vspace*{-1.5cm}
\centerline{\large EUROPEAN ORGANIZATION FOR NUCLEAR RESEARCH (CERN)}
\vspace*{0.5cm}
\noindent
\begin{tabular*}{\linewidth}{lc@{\extracolsep{\fill}}r@{\extracolsep{0pt}}}
\ifthenelse{\boolean{pdflatex}}
{\vspace*{-1.5cm}\mbox{\!\!\!\includegraphics[width=.14\textwidth]{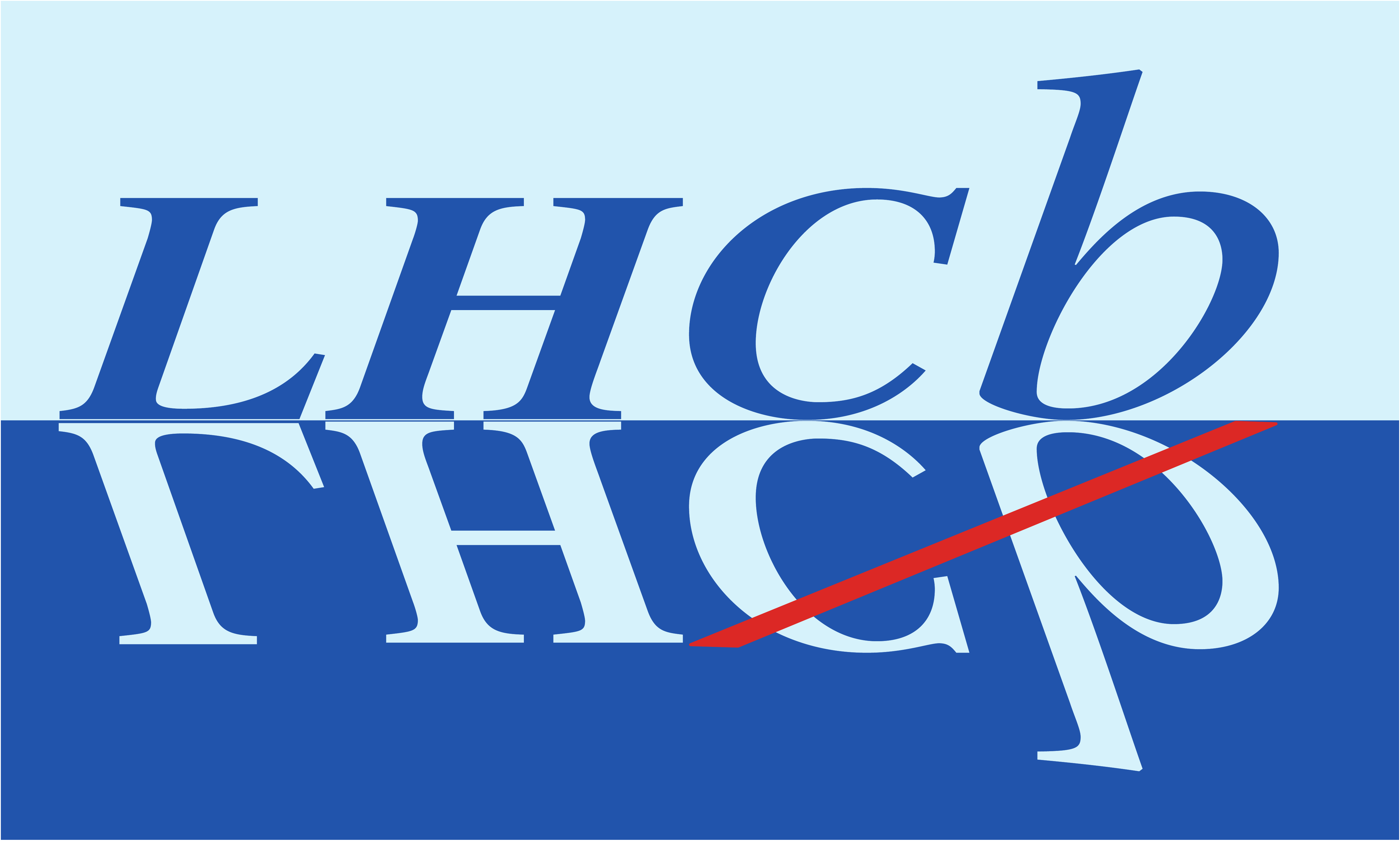}} & &}%
{\vspace*{-1.2cm}\mbox{\!\!\!\includegraphics[width=.12\textwidth]{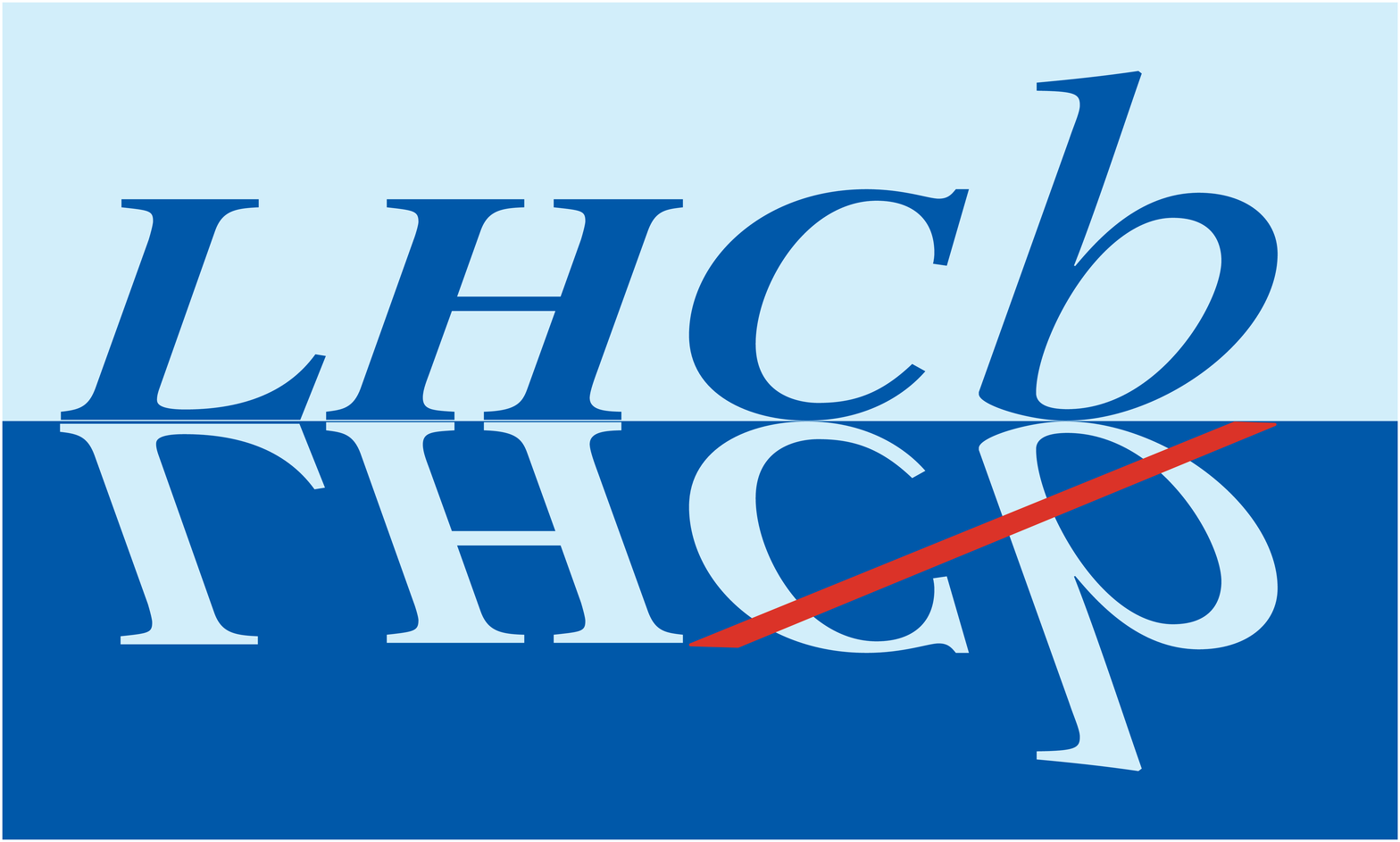}} & &}%
\\
 & & CERN-EP-2020-013 \\  
 & & LHCb-PAPER-2019-045 \\  
%
%
%
   & & February 12,2020 \\ 
%
%
%
\end{tabular*}

\vspace*{2.0cm}

{\normalfont\bfseries\boldmath\huge
\begin{center}
  \papertitle 
\end{center}
}

\vspace*{1.0cm}

\begin{center}
\paperauthors\footnote{Authors are listed at the end of this paper.}
\end{center}

\vspace{\fill}

\begin{abstract}
\noindent
A~new baryon state 
is observed  
in the~$\Lb\pip\pim$~mass 
spectrum 
with high significance 
using a data sample of $\proton\proton$ collisions, collected with the~\lhcb detector 
at centre\nobreakdash-of\nobreakdash-mass energies 
$\sqrt{s}=7, 8$ and 13\tev, corresponding 
to an~integrated luminosity of~9\invfb .
The mass and natural width of the new state are measured to~be 
\begin{eqnarray*}
     m & =  & 6072.3 \pm 2.9 \pm 0.6 \pm 0.2 \mev\,, \\
      \Gamma & = &  72 \pm 11 \pm 2 \mev\,,
\end{eqnarray*}
where the first uncertainty is statistical and the~second systematic.
The~third uncertainty for the~mass is 
due to imprecise knowledge of the~\Lb~baryon mass. 
The~new state  is consistent with the~first radial excitation 
of the~\Lb baryon, the~$\Lambda_\bquark\mathrm{(2S)}^0$~resonance.
Updated measurements of the masses and the upper limits on the natural widths of the previously observed $\Lambdares_{\bquark}(5912)^0$ and 
$\Lambdares_{\bquark}(5920)^0$ states are also reported. 

\end{abstract}

\vspace*{1.0cm}

\begin{center}
  Published in JHEP 06\,(2020) 136.
\end{center}

\vspace{\fill}

{\footnotesize 
\centerline{\copyright~\papercopyright, \href{\paperlicenceurl}{\paperlicence}.}}
\vspace*{2mm}

\end{titlepage}


\newpage
\setcounter{page}{2}
\mbox{~}
%

%
%
%

\cleardoublepage


\renewcommand{\thefootnote}{\arabic{footnote}}
\setcounter{footnote}{0}



\pagestyle{plain} 
\setcounter{page}{1}
\pagenumbering{arabic}


%

\section{Introduction}

The~constituent quark model~\cite{GellMann:1964nj, Zweig:352337,*Zweig:570209} is very successful in describing and classifying the~known hadrons based on their quantum numbers~\cite{PDG2018}. 
However, quantum chromodynamcs that lies
in the~origin of the~quark model, 
being a~nonperturbative theory, does not 
predict  hadron properties, 
namely masses and decay widths,
from first principles. 
Alternative theoretical approaches
are developed, such as 
heavy quark effective theory or lattice calculations.
These approaches require verification with experiment in 
various regimes, \eg testing the~agreement with data 
for hadrons with different quark content and quantum numbers. 
Baryons, containing a~beauty quark form a~particular family of hadrons, 
where the experimental data are still scarce.


Excited beauty baryons with two light quarks\,and quark content $\bquark\quark\quark'$, where $\quark, \quark' = \uquark, \dquark$, have been studied experimentally at the~Tevatron and the~LHC. 
The~family of these baryons consists of the~\Lb~isosinglet 
and the~$\Sigma_{\bquark}$ and $\Sigma_{\bquark}^{\ast}$~isotriplet states. The lightest charged $\Sigmares^{(\ast)\pm}_{\bquark}$ baryons have been observed by the CDF collaboration~\cite{Aaltonen:2007ar, CDF:2011ac} in the $\Lb\pipm$ spectrum. 
The~measurement of the~masses and widths of those states was updated by the~LHCb collaboration 
and the~heavier $\Sigma_{\bquark}(6097)^{\pm}$ states were discovered~\cite{LHCb-PAPER-2018-032}. 

The spectrum of excited beauty baryons decaying to the $\Lb\pip\pim$ final state near threshold has been studied by the~\lhcb collaboration using a data sample collected in 2011, which resulted in the~discovery of two narrow states~\cite{LHCb-PAPER-2012-012}, denoted $\lboneplow$ and $\lbonephigh$.
The~most likely interpretation of these states is that they are a doublet of first orbital excitations in the~\Lb system, with quantum numbers 
${\mathrm{J}}^{\mathrm{P}}=\tfrac{1}{2}^-$ and $\tfrac{3}{2}^-$, respectively. 
The~heavier of these states was later confirmed by the~\cdf collaboration~\cite{Aaltonen:2013tta}. 
A~doublet of narrow states, $\lbonedlow$ and $\lbonedhigh$, was also observed by LHCb collaboration~\cite{LHCb-PAPER-2019-025}.
The~measured  masses and widths of these states are 
compatible with the~expectations 
for the~$\Lambda_{\bquark}\mathrm{(1D)}^0$~doublet~\cite{Chen:2019ywy,Yang:2019cvw,Wang:2019uaj,Liang:2019aag}.
Recently, the~CMS collaboration reported 
an~evidence  for a~broad excess of events
in the~$\Lb\pip\pim$~mass spectrum 
in the~region of~\mbox{$6040-6100\mev$}
corresponding to a~statistical significance of four standard
deviations~\cite{Sirunyan:2020gtz}.\footnote{Natural units are  used through the paper with $c=\hbar=1$.}
The~existence of additional states in the~\lambdabpipi spectrum is predicted by the~quark model~\cite{Capstick:1986bm, Roberts:2007ni,Ebert:2011kk}, notably, in the~region between 
the~established narrow doublet states, with masses around 6.1\gev. 
Quark-model 
predictions for the~masses
of the~lightest $\Lambda_{\bquark}$ and $\Sigma_{\bquark}^{(\ast)}$
states are shown in~Table~\ref{tab:mass_predictions}.

\begin{table}[tb]
  \centering
  \caption{Quark-model 
    predictions for the masses
    of the lightest $\Lambda_{\bquark}$ and $\Sigma_{\bquark}^{(\ast)}$
    states\,(in \mev).}
  \label{tab:mass_predictions}
  \begin{tabular*}{0.95\textwidth}{@{\hspace{3mm}}l@{\extracolsep{\fill}}cccccc@{\hspace{2mm}}}
    Baryon & State & ${\mathrm{J}}^{\mathrm{P}}$  
    & Ref.~\cite{Capstick:1986bm}
    & Ref.~\cite{Roberts:2007ni}
    & Ref.~\cite{Ebert:2011kk} 
    & Ref.~\cite{Chen:2014nyo} 
    \\[1mm]
    \hline 
    \\[-4mm]
    \multirow{6}{*}{$\Lb$} 
    &                  1S & $\tfrac{1}{2}^+$
    & 5585 
    & 5612 
    & 5620 
    & 5619 
    \\[1mm]
    \cline{2-7}
    \\[-4mm]
    & \multirow{2}{*}{1P} &  $\tfrac{1}{2}^-$ 
    & 5912 
    & 5939 
    & 5930 
    & 5911 
    \\
    &                     &  $\tfrac{3}{2}^-$ 
    & 5920 
    & 5941 
    & 5942 
    & 5920 
    \\[1mm]
    \cline{2-7}
    \\[-4mm]
    &                  2S &  $\tfrac{1}{2}^+$ 
    & 6045 
    & 6107 
    & 6089 
    & 
    \\[1mm]
    \cline{2-7}
    \\[-4mm]
    & \multirow{2}{*}{1D} &  $\tfrac{3}{2}^+$ 
    & 6145 
    & 6181 
    & 6190 
    & 6147 
    \\
    &                     &  $\tfrac{5}{2}^+$ 
    & 6165 
    & 6183
    & 6196 
    & 6153 
    \\[1mm]
    \hline 
    \\[-4mm]
    \multirow{7}{*}{$\Sigma_b^{(\ast)0}$} 
    & \multirow{2}{*}{1S} &  $\tfrac{1}{2}^+$ 
    & 5795 
    & 5833 
    & 5800 
    & 
    \\
    &                     &  $\tfrac{3}{2}^+$ 
    & 5805 
    & 5858 
    & 5834 
    & 
    \\[1mm]
    \cline{2-7}
    \\[-4mm]
    & \multirow{3}{*}{1P} &  $\tfrac{1}{2}^-$  
    & 6070 
    & 6099 
    & 6101
    & 
    \\
    &                     &  $\tfrac{3}{2}^-$ 
    & 6070 
    & 6101 
    & 6096 
    \\
    &                     &  $\tfrac{5}{2}^-$
    & 6090 
    & 6172 
    & 6084 
    & 
    \\[1mm]
    \cline{2-7}
    \\[-4mm]
    & \multirow{2}{*}{2S} &  $\tfrac{1}{2}^+$  
    & 6200 
    & 6294 
    & 6213 
    & 
    \\
    &                     &  $\tfrac{3}{2}^+$ 
    & 6250 
    & 6308 
    & 6226 
    & 
  \end{tabular*}
\end{table}

This paper reports the observation of a~new structure in the~\Lb\pip\pim mass spectrum, as well as updated measurements of 
the~masses and widths of the~$\lboneplow$ and $\lbonephigh$ states with improved precision. 
The~analysis uses 
$\proton\proton$ collision data 
recorded by LHCb in~\mbox{2011--2018} 
at centre\nobreakdash-of\nobreakdash-mass energies of 
7, 8 and 13\tev, 
corresponding to an~integrated 
luminosity of~1, 2 and 
6\invfb, respectively.

\section{The LHCb detector}
The \lhcb detector~\cite{Alves:2008zz,LHCb-DP-2014-002} is a~single\nobreakdash-arm forward
spectrometer covering the~\mbox{pseudorapidity} range~\mbox{$2<\eta <5$},
designed for the study of particles containing \bquark~or \cquark~quarks.
The detector includes a~high\nobreakdash-precision tracking system
consisting of a~silicon\nobreakdash-strip vertex detector surrounding 
the~$\proton\proton$~interaction region~\cite{LHCb-DP-2014-001}, 
a~large\nobreakdash-area silicon\nobreakdash-strip detector located
upstream of a dipole magnet with a bending power of about
$4{\mathrm{\,Tm}}$, and three stations of silicon-strip detectors and straw
drift tubes~\cite{LHCb-DP-2013-003,LHCb-DP-2017-001} placed downstream of the~magnet.
The~tracking system provides a measurement of the~momentum, \ptot, of charged particles with
a~relative uncertainty that varies from 0.5\% at 
low momentum
to 1.0\% at 200\gev.
The~momentum scale of the~tracking system is calibrated using samples of 
$\decay{\jpsi}{\mumu}$
and $\decay{\Bu}{\jpsi\Kp}$ decays collected concurrently with 
the~data sample
used for this 
analysis\mbox{\cite{LHCb-PAPER-2012-048,LHCb-PAPER-2013-011}}.
The~relative accuracy of this procedure is estimated to be
$3 \times 10^{-4}$ using samples of other fully reconstructed
$\bquark$-hadron, $\KS$, and narrow \mbox{$\PUpsilon\mathrm{(1S)}$}~resonance decays.
Different types of charged hadrons are distinguished by 
the~particle 
identification\,(PID) system using information
from two ring-imaging Cherenkov detectors~\cite{LHCb-DP-2012-003}.
Muons are identified by a system composed of alternating layers of iron and multiwire proportional chambers~\cite{LHCb-DP-2012-002}.

The~online event selection is performed by a~trigger~\cite{LHCb-DP-2012-004}
which consists of a~hardware stage, based on information 
from the calorimeter and muon systems, followed by a software stage, 
which applies a full event reconstruction.
At the hardware trigger stage, 
events are required to have a~muon with high transverse momentum,~\pt,
or a~pair of opposite\nobreakdash-sign muons with a~requirement
on the~product of muon transverse momenta, 
or a~hadron, photon or electron with 
high transverse energy in the~calorimeters. 
The~software trigger requires a~two-, three- or four\nobreakdash-track
secondary vertex with at least one charged particle with a~large \pt
and inconsistent with originating from any 
reconstructed primary
$\proton\proton$~collision vertex\,(PV)~\cite{BBDT,LHCb-PROC-2015-018} or  
two muons of opposite charge forming 
a~good\nobreakdash-quality secondary vertex with a~mass in excess 
of~2.7\gev.

Simulation is required to model the effects of the detector acceptance, resolution, and selection requirements. In the simulation, $\proton\proton$ collisions are generated using
\pythia~\cite{Sjostrand:2006za} with a specific \lhcb configuration~\cite{LHCb-PROC-2010-056}. Decays of unstable particles
are described by \evtgen~\cite{Lange:2001uf}, in which final-state
radiation is generated using \photos~\cite{Golonka:2005pn}.
The~interaction of the~generated particles with the detector, and its response,
are implemented using the \geant toolkit~\cite{Allison:2006ve, *Agostinelli:2002hh} as described in Ref.~\cite{LHCb-PROC-2011-006}. 

\section{Event selection}\label{sec:selection}

The~\Lb~candidates are reconstructed in the~$\decay{\Lb}{\Lc\pim}$
and the~$\decay{\Lb}{\jpsi\proton\Km}$~decays.\footnote{Inclusion of 
charge-conjugate states is implied throughout this paper.}
The~selection of the~\Lb~candidates is similar to that used in  Ref.~\cite{LHCb-PAPER-2019-025}.
All~charged final\nobreakdash-state particles are required to be positively identified by the PID systems. 
To~reduce the~background from random combinations of tracks, 
only the~tracks with large impact parameter with respect 
to all PVs in the event are used.
The~$\Lc$ candidates are
reconstructed in the~$\proton\Km\pip$ final state.
The~$\lbjpsipk$ candidates are created by combining 
the~$\jpsi$~candidates formed of  $\mumu$~pairs with kaon and proton tracks.  
The~masses of the~$\Lc$ and $\jpsi$~candidates are required to be consistent 
with the~known values of the~masses of the~respective states~\cite{PDG2018}
and the~$\Lb$~candidate is required to have a~good\nobreakdash-quality 
vertex  significantly displaced from all~PVs. 

\begin{figure}[tb]
  \setlength{\unitlength}{1mm}
  \centering
  \begin{picture}(150,60)
    %
    \put(  0, 0){ 
      \includegraphics*[width=75mm,height=60mm,%
      ]{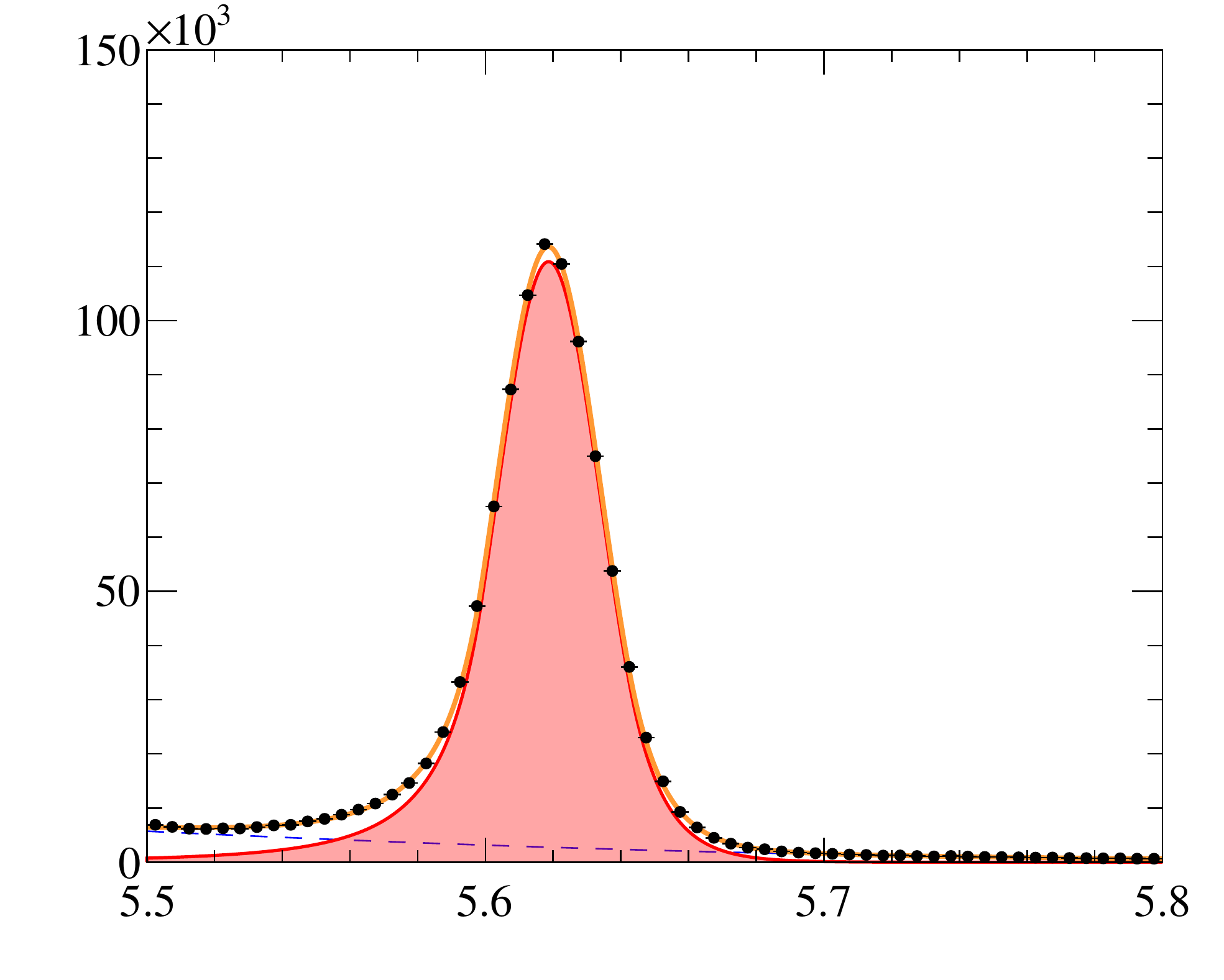}
    } 
    \put( 75, 0){ 
      \includegraphics*[width=75mm,height=60mm,%
      ]{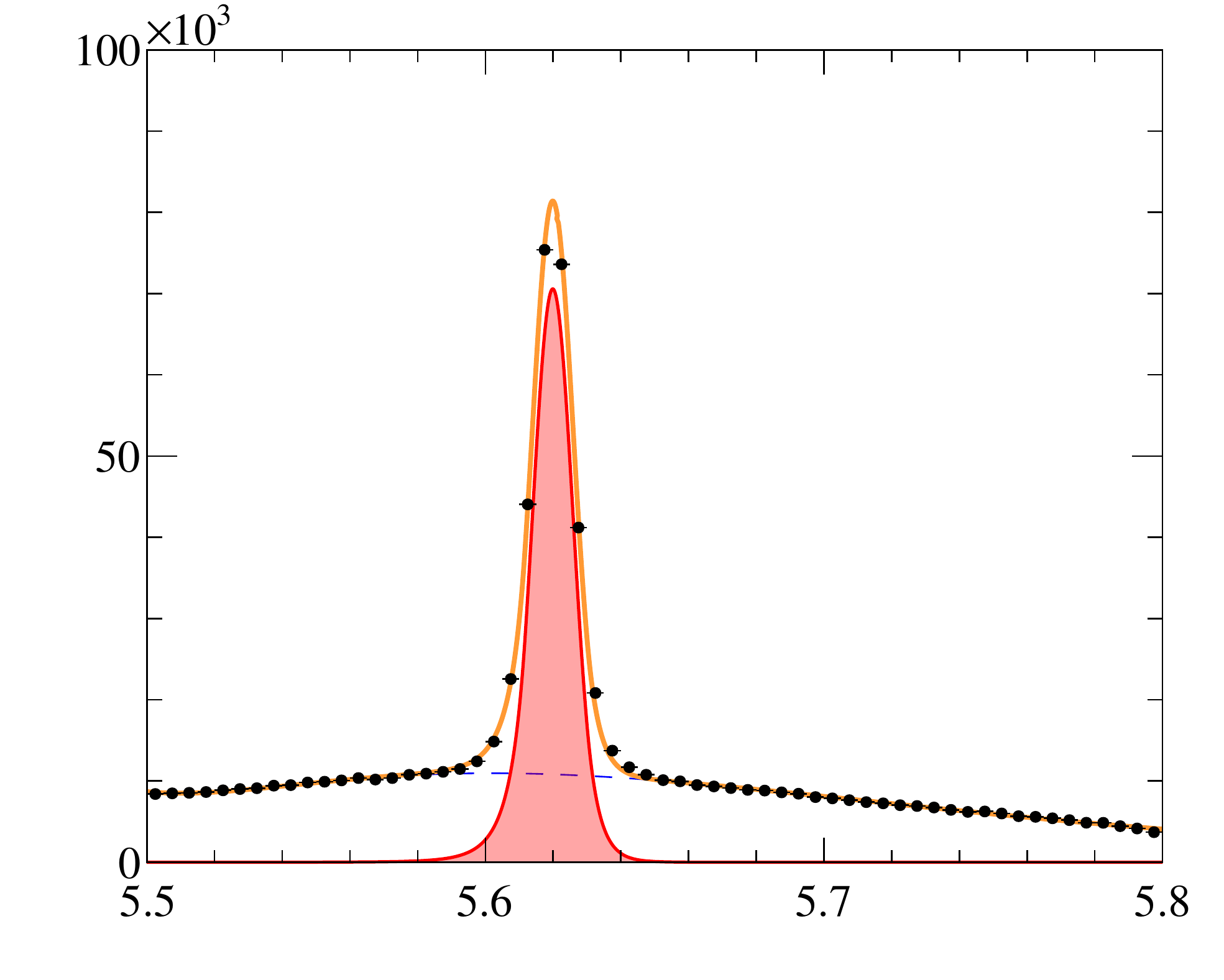}
    } 
    \put( 1,20){\begin{sideways}Candidates/$\left(5\mev\right)$
      \end{sideways}}
    \put(76,20){\begin{sideways}Candidates/$\left(5\mev\right)$
      \end{sideways}}
     
    \put(35 ,0){$m_{\Lc\pim}$}
    \put(110,0){$m_{\jpsi\proton\Km}$}
    \put( 63,0){$\left[\!\gev\right]$}
    \put(138,0){$\left[\!\gev\right]$}
    \put( 45,47){$\begin{array}{r}\lhcb \\ \decay{\Lb}{\Lc\pim}\end{array}$}
    \put(117,47){$\begin{array}{r}\lhcb \\ \decay{\Lb}{\jpsi\proton\Km}\end{array}$}
    \put( 14,52) {\begin{tikzpicture}[x=1mm,y=1mm]\filldraw[fill=red!35!white,draw=red,thick]  (0,0) rectangle (6,1.5);\end{tikzpicture} }
    \put( 14,48) {\color[rgb]{0.00,0.00,1.00} {\hdashrule[0.5ex][x]{6mm}{0.3mm}{2.0mm 1.0mm} } } 
    \put( 14,44) {\color[rgb]{1.00,0.65,0.00} {\hdashrule[0.5ex][x]{6mm}{0.8mm}{1.0mm 0.0mm} } } 
    \put( 22,52){\small{signal}}
    \put( 22,48){\small{background}}
    \put( 22,44){\small{total}}
  \end{picture}
  \caption { \small
    Mass distributions for selected 
        (left)~\mbox{$\decay{\Lb}{\Lc\pim}$}
        and 
        (right)~\mbox{$\decay{\Lb}{\jpsi\proton\Km}$}
    candidates after BDT selection.
    A fit, composed of a~sum of 
    a~double-sided Crystal Ball function~\cite{Skwarnicki:1986xj} 
    and a~smooth background component, is overlaid. 
  }
  \label{fig:Lb}
\end{figure}

Further suppression of the background is achieved by using 
a~boosted decision tree\,(BDT) classifier~\cite{Breiman,AdaBoost} implemented 
in the~TMVA
toolkit~\cite{Hocker:2007ht,*TMVA4}. 
Two~separate BDTs are used for the~$\decay{\Lb}{\Lc\pim}$
and $\decay{\Lb}{\jpsi\proton\Km}$~selections.
The~multivariate estimators are based on 
the~kinematic properties, the reconstructed lifetime   and 
vertex quality  of the~\Lb~candidate
and on variables describing the~overall 
consistency of the~selected candidates 
with the~decay chain obtained
from the~kinematic fit described below~\cite{Hulsbergen:2005pu}.
In~addition, 
the~reconstructed lifetime and vertex quality 
of the~\mbox{$\decay{\Lc}{\proton\Km\pip}$}~candidate is used for~the~\mbox{$\decay{\Lb}{\Lc\pim}$} decay.
The~PID quality, transverse momentum and \mbox{pseudorapidity} 
of the~proton and kaon~candidates\,(for $\decay{\Lb}{\jpsi\proton\Km}$)
or \pim~candidate\,(for $\decay{\Lb}{\Lc\pim}$) are also used. 
The~BDT is trained using data, where the~signal 
sample is obtained 
by subtracting the~background using the~\sPlot technique~\cite{Pivk:2004ty}, 
and the~background sample is taken from the~range $5.70-5.85\gev$ in 
the~$\decay{\Lb}{\Lc\pim}$
and $\decay{\Lb}{\jpsi\proton\Km}$ mass distributions. 
A~$k$-fold cross\nobreakdash-validation technique 
is used to avoid introducing a~bias in the~evaluation~\cite{geisser1993predictive}.  
A~kinematic fit~\cite{Hulsbergen:2005pu} is performed in order to improve the \Lb~mass resolution.
The~momenta of the~particles in the~full decay chain are recomputed
by constraining 
 the~\Lc or~\jpsi~mass  
to their~known values~\cite{PDG2018}
and the~\Lb~baryon to originate 
from the~associated PV.
The~mass distributions for the~selected 
\mbox{$\decay{\Lb}{\Lc\pim}$} and 
\mbox{$\decay{\Lb}{\jpsi\proton\Km}$}~candidates
are shown in Fig.~\ref{fig:Lb}.
The~\Lb~signal yield 
is
$\left(937.9\pm1.6\right)\times 10^3$
and 
$\left(223.0\pm0.6\right)\times 10^3$
for 
\mbox{$\decay{\Lb}{\Lc\pim}$} and 
\mbox{$\decay{\Lb}{\jpsi\proton\Km}$}~decays, respectively.

Selected  $\decay{\Lb}{\Lc\pim}\,(\decay{\Lb}{\jpsi\proton\Km})$ 
candidates with mass within $\pm50$\,($20$)\mev
from the~known \Lb~mass are 
combined with pairs of 
opposite and same\nobreakdash-sign pion tracks.
To~reduce the~large combinatorial background, 
four separate BDT classifiers are trained for the~$\lblcpi$ and 
$\lbjpsipk$ samples in 
the~high\nobreakdash-mass\,\mbox{($m_{\Lb\Ppi\Ppi}<6.35\gev$)} 
and the~low\nobreakdash-mass\,\mbox{($m_{\Lb\Ppi\Ppi}<5.95\gev$)} regions. 
The~BDTs exploit the~vertex quality, $\chisq_{\rm vtx}$,  of 
the~$\Lb\Ppi\Ppi$ combination, its transverse momentum, 
the~$\pt$ of the~$\Ppi\Ppi$ pair, 
the~$\pt$ of each pion, as well as their PID 
and track\nobreakdash-reconstruction\nobreakdash-quality variables.
For the~high-mass region, 
the~$\pt$ of the~dipion~system is required to exceed 250\mev.
Simulated samples of excited \Lb~baryons
decaying into the~$\Lb\pip\pim$~final state are used as 
signal training samples, 
while the~background training sample is taken from the~same\nobreakdash-sign $\lambdabpipiSS$ 
combinations in data. For~the~low\nobreakdash-mass region, 
simulated samples of $\Lambda_{\bquark}\mathrm{(5912)}^0$ and 
$\Lambda_{\bquark}\mathrm{(5920)}^0$~signal decays are used, 
while for 
the~high\nobreakdash-mass region 
the~simulated sample consists of decays of a~narrow  state with mass of 6.15\gev and natural 
width of 7\mev, and a~broad state  with mass of 6.08\gev and 
natural width of 60\mev.
A~$k$-fold cross\nobreakdash-validation 
technique 
is used for training.  
A~figure of merit $\varepsilon/(\tfrac{5}{2}+\sqrt{B})$~\cite{Punzi:2003bu}
is used to optimise the~requirement on the~BDT estimator.
The~$\Lb\Ppi\Ppi$~mass resolution
is improved by a~kinematic fit~\cite{Hulsbergen:2005pu}
constraining the~mass of 
the~$\proton\Km\pip$ and 
$\mumu$~combinations to 
the~known masses of the~\Lc~baryon and \jpsi~meson, 
respectively~\cite{PDG2018}.
The~mass  of the~\Lb~baryon in the~fit 
is constrained to the central value
of \mbox{$m_{\Lb}=5619.62\pm0.16\pm0.13\mev$}~\cite{LHCb-PAPER-2017-011}.
It is also required that 
the~momentum vector of the~\Lb~candidate 
and the momenta of both pions points back to 
the~associated $\proton\proton$~interaction vertex.

\section{Analysis of the~high\nobreakdash-mass region} 

\begin{figure}[t]
  \setlength{\unitlength}{1mm}
  \centering
  \begin{picture}(150,150)
  \definecolor{root8}{rgb}{0.35,0.83,0.33}
    \put(  0, 0){ 
      \includegraphics*[height=150mm,width=150mm,%
      ]{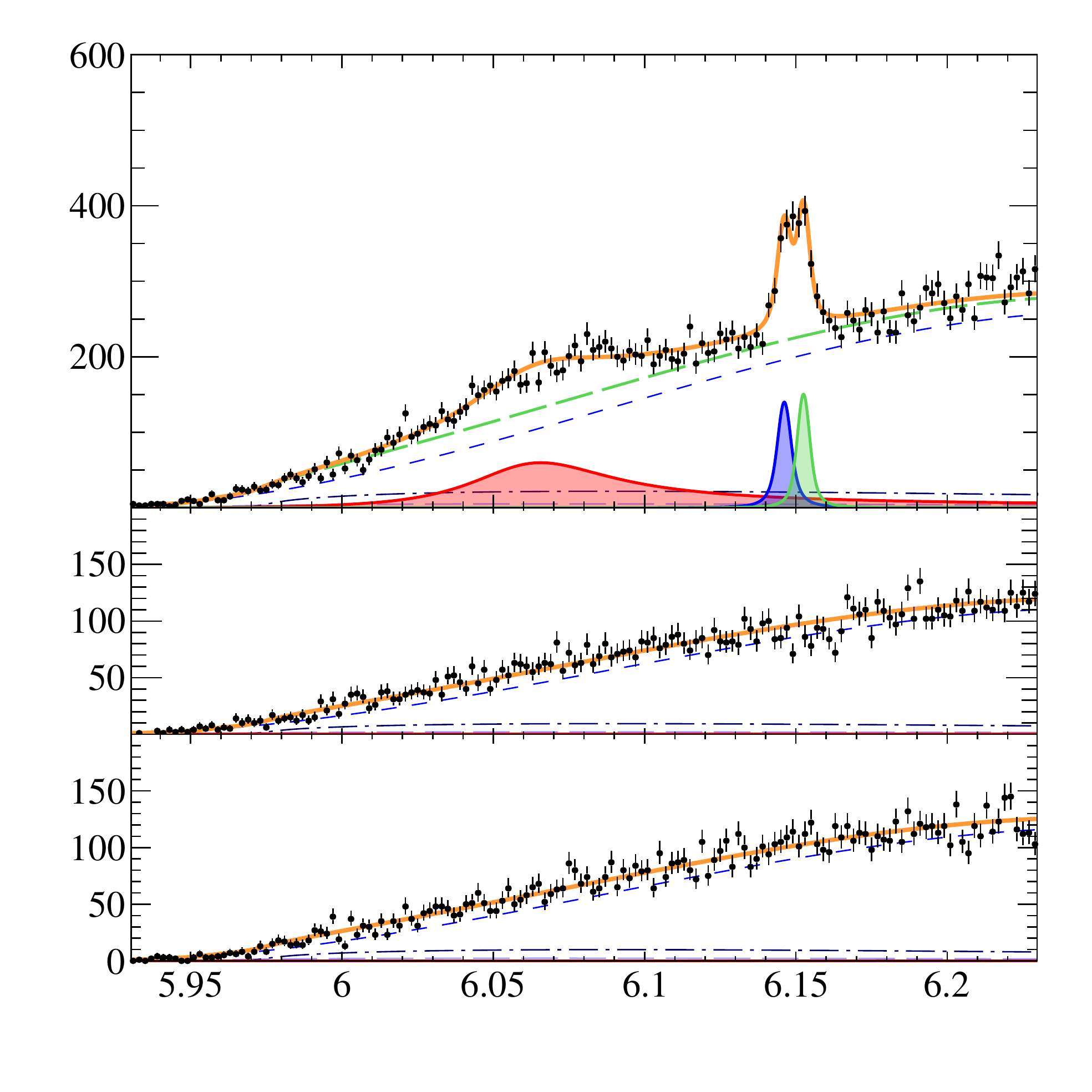}
    }  
    \put(25,130) {\begin{tikzpicture}[x=1mm,y=1mm]\filldraw[fill=red!35!white,draw=red,thick]  (0,0) rectangle (12,1.5);\end{tikzpicture} }
    \put(25,126) {\begin{tikzpicture}[x=1mm,y=1mm]\filldraw[fill=blue!35!white,draw=blue,thick]  (0,0) rectangle (12,1.5);\end{tikzpicture} }
    \put(25,122) {\begin{tikzpicture}[x=1mm,y=1mm]\filldraw[fill=root8!35!white,draw=root8,thick]  (0,0) rectangle (12,1.5);\end{tikzpicture} }
    \put(25,118) {\color[rgb]{0.60,0.60,1.00}{\hdashrule[0.5ex][x]{12mm}{0.3mm}{3.5mm 0.5mm} } } 
    \put(25,114) {\color[rgb]{0.00,0.00,0.40}{\hdashrule[0.5ex][x]{12mm}{0.3mm}{2.5mm 0.5mm 0.5mm 0.5mm} } } 
    \put(25,110) {\color[rgb]{0.00,0.00,1.00} {\hdashrule[0.5ex][x]{12mm}{0.5mm}{1.5mm 1.5mm} } } 
    \put(25,106) {\color[rgb]{0.35,0.83,0.33} {\hdashrule[0.5ex][x]{12mm}{1.0mm}{5.0mm 1.0mm} } } 
    \put(25,102) {\color[rgb]{1.00,0.65,0.00} {\hdashrule[0.5ex][x]{12mm}{1.0mm}{1.0mm 0.0mm} } } 
    \put(40,130) {\small{$\Lambda_{\bquark}^{\ast\ast0}$}}
    \put(40,126) {\small{$\Lambda_{\bquark}\mathrm{(6146)}^0$}}
    \put(40,122) {\small{$\Lambda_{\bquark}\mathrm{(6152)}^0$}}
    \put(40,118) {\small{$\Sigma_{\bquark}\Ppi$}}
    \put(40,114) {\small{$\Sigma_{\bquark}^{\ast}\Ppi$}}
    \put(40,110) {\small{comb. background}}
    \put(40,106) {\small{total background}}
    \put(40,102) {\small{total}}
    \put(5 ,73){\large\begin{sideways}Candidates/$\left(2\mev\right)$\end{sideways}}
    \put(75, 3){\large{$m_{\Lb\Ppi\Ppi}$}}\put(132, 3) {\large{$\left[\!\gev\right]$}}
    \put(55,135){\large{$\decay{\Lb}{\Lc\pim}$}}
    \put(25,135){\large{$\Lb\pip\pim$}}
    \put(25, 73){\large{$\Lb\pip\pip$}}
    \put(25, 42){\large{$\Lb\pim\pim$}}
    \put(125,135){\large\lhcb}
  \end{picture}
  \caption { \small
    Mass spectra of selected 
    (top)~$\Lb\pip\pim$,
    (middle)~$\Lb\pip\pip$ and 
    (bottom)~$\Lb\pim\pim$~combinations for 
    the~$\decay{\Lb}{\Lc\pim}$~sample.
    A~simultaneous fit, described in the~text, is superimposed.
  }
  \label{fig:fit_lcpi}
\end{figure}

\begin{figure}[htb]
  \setlength{\unitlength}{1mm}
  \centering
  \begin{picture}(150,150)
  \definecolor{root8}{rgb}{0.35,0.83,0.33}
    \put(  0, 0){ 
      \includegraphics*[height=150mm,width=150mm,%
      ]{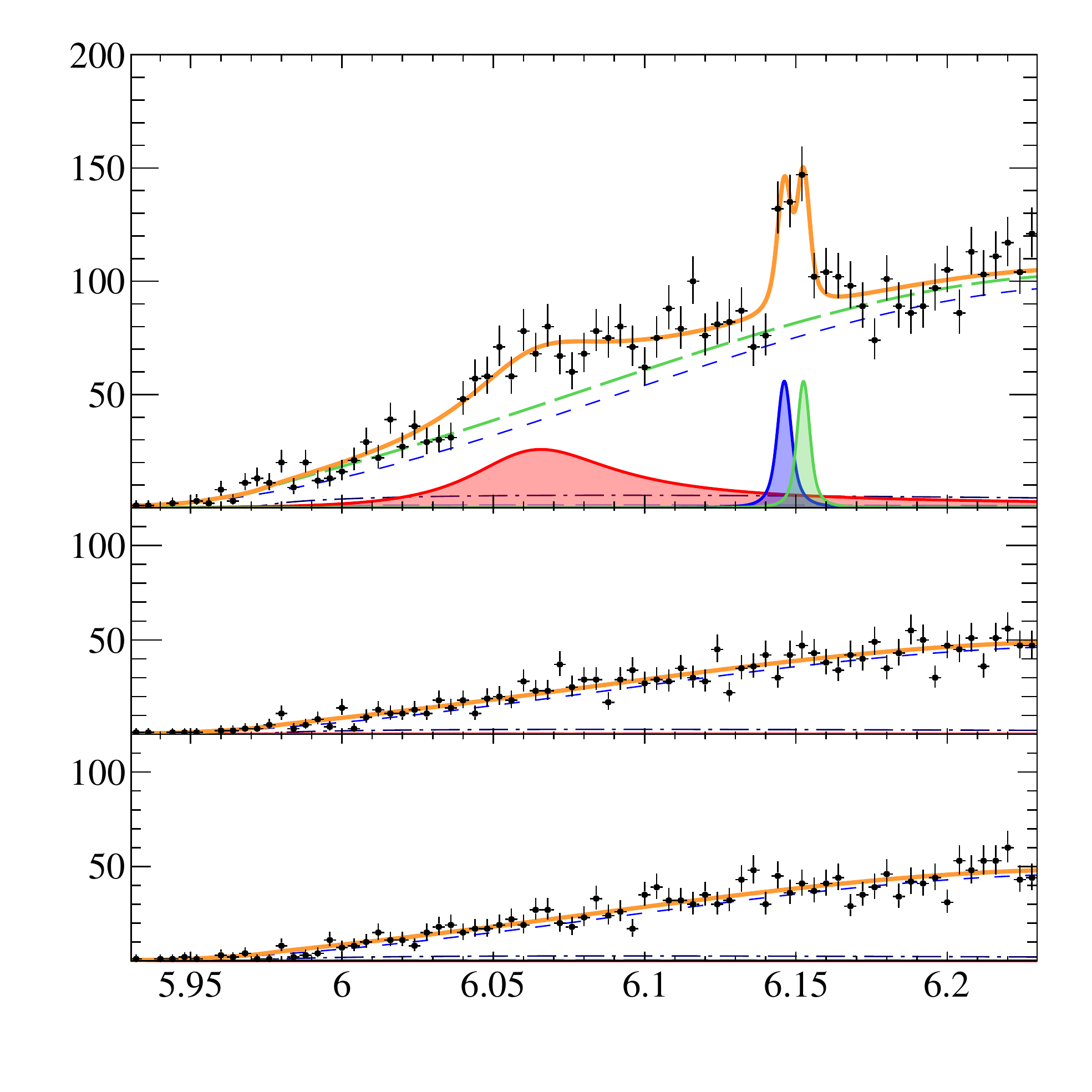}
    }  
    \put(25,130) {\begin{tikzpicture}[x=1mm,y=1mm]\filldraw[fill=red!35!white,draw=red,thick]  (0,0) rectangle (12,1.5);\end{tikzpicture} }
    \put(25,126) {\begin{tikzpicture}[x=1mm,y=1mm]\filldraw[fill=blue!35!white,draw=blue,thick]  (0,0) rectangle (12,1.5);\end{tikzpicture} }
    \put(25,122) {\begin{tikzpicture}[x=1mm,y=1mm]\filldraw[fill=root8!35!white,draw=root8,thick]  (0,0) rectangle (12,1.5);\end{tikzpicture} }
    \put(25,118) {\color[rgb]{0.60,0.60,1.00}{\hdashrule[0.5ex][x]{12mm}{0.3mm}{3.5mm 0.5mm} } } 
    \put(25,114) {\color[rgb]{0.00,0.00,0.40}{\hdashrule[0.5ex][x]{12mm}{0.3mm}{2.5mm 0.5mm 0.5mm 0.5mm} } } 
    \put(25,110) {\color[rgb]{0.00,0.00,1.00} {\hdashrule[0.5ex][x]{12mm}{0.5mm}{1.5mm 1.5mm} } } 
    \put(25,106) {\color[rgb]{0.35,0.83,0.33} {\hdashrule[0.5ex][x]{12mm}{1.0mm}{5.0mm 1.0mm} } } 
    \put(25,102) {\color[rgb]{1.00,0.65,0.00} {\hdashrule[0.5ex][x]{12mm}{1.0mm}{1.0mm 0.0mm} } } 
    \put(40,130) {\small{$\Lambda_{\bquark}^{\ast\ast0}$}}
    \put(40,126) {\small{$\Lambda_{\bquark}\mathrm{(6146)}^0$}}
    \put(40,122) {\small{$\Lambda_{\bquark}\mathrm{(6152)}^0$}}
    \put(40,118) {\small{$\Sigma_{\bquark}\Ppi$}}
    \put(40,114) {\small{$\Sigma_{\bquark}^{\ast}\Ppi$}}
    \put(40,110) {\small{comb. background}}
    \put(40,106) {\small{total background}}
    \put(40,102) {\small{total}}
    \put(5 ,73){\large\begin{sideways}Candidates/$\left(4\mev\right)$\end{sideways}}
    \put(75, 3){\large{$m_{\Lb\Ppi\Ppi}$}}\put(132, 3) {\large{$\left[\!\gev\right]$}}
    \put(55,135){\large{$\decay{\Lb}{\jpsi\proton\Km}$}}
    \put(25,135){\large{$\Lb\pip\pim$}}
    \put(25, 73){\large{$\Lb\pip\pip$}}
    \put(25, 42){\large{$\Lb\pim\pim$}}
    \put(125,135){\large\lhcb}
  \end{picture}
  \caption { \small
  Mass spectra of selected 
    (top)~$\Lb\pip\pim$,
    (middle)~$\Lb\pip\pip$ and 
    (bottom)~$\Lb\pim\pim$~combinations for
    the~$\decay{\Lb}{\jpsi\proton\Km}$~sample.
    A~simultaneous fit, described in the~text, is superimposed.
   }
  \label{fig:fit_psipk}
\end{figure}

The distributions of the $\lambdabpipi$ and $\lambdabpipiSS$ masses
in the~range \mbox{$5.93<m_{\Lb\Ppi\Ppi}<6.23\gev$} 
for the~$\lblcpi$ sample with the~high\nobreakdash-mass BDT selection applied are shown in 
Fig.~\ref{fig:fit_lcpi}. 
The~distributions of the same-sign $\lambdabpipiSS$
combinations are dominated by random combinations of 
a~\Lb baryon and two pions.
The~$\lambdabpipi$ spectrum features
the~contributions of two narrow $\lbonedlow$ and $\lbonedhigh$
states as well as a~broad structure just below~6.1\gev
in addition to the~smooth background.
This~new structure is referred to as $\lbstarstar$ hereafter. 
Figure~\ref{fig:fit_psipk} shows the~same distributions 
for the~$\lbjpsipk$
sample, where the~same features are visible.

A~simultaneous binned  maximum\nobreakdash-likelihood fit
with a~bin width of 200\kev
is performed to 
the~six 
distributions shown in Figs.~\ref{fig:fit_lcpi} and \ref{fig:fit_psipk}
in order to determine the~properties of the~resonant shapes.
Both signal and background  $\Lb\Ppi\Ppi$~combinations
could include contributions from 
intermediate $\Sigma_{\bquark}^{\pm}$ and  
$\Sigma_{\bquark}^{\ast\pm}$~states. 
The~fitting function for the~\lambdabpipi spectra is the~sum 
of five components: 
a~combinatorial background, 
the~two components corresponding to 
the~combinations of 
\mbox{$\decay{\Sigma_{\bquark}^{\pm}}{\Lb\Ppi^{\pm}}$} 
and 
\mbox{$\decay{\Sigma_{\bquark}^{\ast\pm}}{\Lb\Ppi^{\pm}}$} 
with the~addition of a~pion from the~rest of the~event,
and three resonant contributions for the $\lbonedlow$, 
$\lbonedhigh$ and  $\lbstarstar$~states. 
The same-sign $\Lb\Ppi^{\pm}\Ppi^{\pm}$ spectra are fitted with 
a~function that contains only 
the~combinatorial, 
$\Sigma_{\bquark}^{\pm}\Ppi^{\pm}$, and 
$\Sigma_{\bquark}^{\ast\pm}\Ppi^{\pm}$ components.

The combinatorial background is parameterised with a~positive,
increasing third\nobreakdash-order polynomial function, 
whose~coefficients are left free to vary in the~fit. 
The~$\Sigma_{\bquark}^{\pm}\Ppi$ and 
$\Sigma_{\bquark}^{\ast\pm}\Ppi$ components are
described by the~product of a~two\nobreakdash-body phase-space 
function and an~exponential function, accounting for the~finite width of
the~$\Sigma_{\bquark}^{(\ast)}$~states.
The~exponential factor is determined from the~fit to 
the~background\nobreakdash-subtracted 
$\Sigmares_{\bquark}^{(*)\pm}\Ppi$~mass distributions
in the~\mbox{$6.16<m_{\Lb\Ppi\Ppi}<6.40\gev$} range.
The~shapes of the~$\Sigmares^{(*)\pm}_{\bquark}\Ppi$~components are taken 
to be the~same in all spectra.
The~combinatorial background shape is fixed to be the~same in the~opposite\nobreakdash-sign \lambdabpipi and same\nobreakdash-sign \lambdabpipiSS spectra, but is allowed to differ for the~\lblcpi and \lbjpsipk samples. The~yields of all background components 
are left free to vary in the~fit. 
    A~good description of
    both the~$\Lb\pip\pip$
    and $\Lb\pim\pim$~mass spectra
    supports the~chosen background model.

The~narrow 
$\Lambda_{\bquark}\mathrm{(6146)}^0$ and 
$\Lambda_{\bquark}\mathrm{(6152)}^0$~components are parameterised 
using 
relativistic Breit--Wigner distributions convolved with the experimental resolution.
The~detector resolution function is described by 
the~sum of two Gaussian functions with zero mean 
and parameters  fixed from simulation.
The~obtained effective resolution increases 
from 0.5\mev 
to 1.7\mev when the~$\Lb\pip\pim$~mass grows from 
the~mass of the~$\Lambda_{\bquark}\mathrm{(5912)}^0$~state to 
that  of the~$\Lambda_{\bquark}\mathrm{(6152)}^0$~state.
The~masses and widths of the~\lbonedlow and \lbonedhigh states are fixed to the values obtained in Ref.~\cite{LHCb-PAPER-2019-025}.
The~$\lbstarstar$~shape 
as a~function of the~$\Lb\Ppi\Ppi$~mass $m$
is parameterised as
\begin{equation}
  \mathfrak{S}(m|m_0, \Gamma ) \propto
  \dfrac{
    \Gamma \Prho_3(m) 
  }{
    \left( m^2_0 - m^2 \right)^2 + 
    m_0^2\Gamma^2 
    \left( \dfrac{\Prho_3(m)}{\Prho_3(m_0)} \right)^2
  },  \label{eq:signal}
\end{equation}
where $\Prho_3\left(m\right)$ is a~three\nobreakdash-body  phase space
of the~$\Lb\pip\pim$~system
\begin{equation}
  \Prho_3(m)  
  \equiv 
  \frac{\pi^2}{4m^2} \int\limits_{4m_{\Ppi}^2}^{(m-m_{\Lb})^2} \frac{d m^2_{\Ppi\Ppi}}{ m^2_{\Ppi\Ppi}}\,\,
  \uplambda^{1/2}\left( m^2_{\Ppi\Ppi}, m^2, m_{\Lb}^2\right) 
  \uplambda^{1/2}\left( m^2_{\Ppi\Ppi}, m^2_{\Ppi}, m^2_{\Ppi}\right)\,,
\end{equation}
$\uplambda\left(x,y,z\right)$ stands for  a~K\"{a}ll\'{e}n function~\cite{Kallen},
and $m_{\Ppi}$ and $m_{\Lb}$ denote the~known masses of the~charged $\Ppi$~meson
and $\Lb$~baryon, respectively.
The~mass, $m_0$, and width, $\Gamma$, of the~\lbstarstar~state are free parameters of the fit.

The yields of the fit components in the combined fit are reported in Table~\ref{tab:baryon_yields}. 
The~mass difference with respect to the~\Lb~baryon mass and 
the~natural width of the \lbstarstar state are determined to be
\begin{subequations}
  \begin{eqnarray*}
    \Delta m_{\lbstarstar} & = & \phantom{0}452.7 \pm 2.9 \mev\,,\\ 
    \Gamma_{\Lambda_{\bquark}^{\ast\ast}}   & = & \phantom{00}72\phantom{.0} \pm11\phantom{.} \mev\,,
  \end{eqnarray*}\label{eq:fits_both}\end{subequations}
where  uncertainties are statistical only.
The statistical significance of the~$\lbstarstar$~signal 
in $\decay{\Lb}{\Lc\pim}$ and $\decay{\Lb}{\jpsi\proton\Km}$ samples 
is obtained using Wilks' theorem~\cite{Wilks:1938dza} 
and exceeds 14 and 7 standard deviations, respectively. 
    The~ratios of
    the~$\Lambda_{\bquark}^{\ast\ast0}$,
    $\Lambda_{\bquark}\mathrm{(6146)}^0$ and 
    $\Lambda_{\bquark}\mathrm{(6152)}^0$ signal 
    yields
    between
    the~\mbox{$\decay{\Lb}{\Lc\pim}$} and
    \mbox{$\decay{\Lb}{\jpsi\proton\Km}$}~final state
    are larger than the~ratio of their yields 
    reported in Sec.~\ref{sec:selection}.
    This~arises due to the~differece in
    the~\pt~spectra selected by the~trigger
    for these final states which is propagated to 
    the~$\Ppi\Ppi$~reconstruction effects.

\begin{table}[t]
  \centering
  \caption{ \small
    Yields of excited baryons from the 
    simultaneous fit to $\Lb\Ppi\Ppi$~spectra
    with $\decay{\Lb}{\Lc\pim}$ and $\decay{\Lb}{\jpsi\proton\Km}$. 
  } \label{tab:baryon_yields}
  \vspace*{2mm}
  \begin{tabular*}{0.70\textwidth}{@{\hspace{5mm}}l@{\extracolsep{\fill}}cc@{\hspace{5mm}}}
    &   $\decay{\Lb}{\Lc\pim}$ & $\decay{\Lb}{\jpsi\proton\Km}$
    \\[1mm]
    \hline 
    \\[-2mm]
    $\lbstarstar$           & $2570 \pm260$                       &  $550\pm 80$  \\
    $\Lambda_{\bquark}\mathrm{(6146)}^{0}$  & $\phantom{0}520 \pm 50\phantom{0}$ &  $103\pm 22$   \\
    $\Lambda_{\bquark}\mathrm{(6152)}^{0}$  & $\phantom{0}480 \pm 50\phantom{0}$ &  $\phantom{0}90\pm 21$  
  \end{tabular*}
\end{table}

The~earlier analysis of \lbonedlow and \lbonedhigh states~\cite{LHCb-PAPER-2019-025} has shown 
that a~significant fraction of their decays into the~\lambdabpipi final state proceeds via 
the~intermediate $\Sigma_{\bquark}^{\pm}\Ppi^{\mp}$ 
and $\Sigma_{\bquark}^{\ast\pm}\Ppi^{\mp}$ processes. 
Since the~measured mass of 
the~\lbstarstar state is above the~$\Sigmab\Ppi$ threshold, 
one might expect that this state   
decays via intermediate 
~$\Sigma_{\bquark}^{(\ast)\pm}\Ppi^{\mp}$~states  as well. 
However,  performing the~fits to the~$\Sigmares^{(*)}_{\bquark}\Ppi$~mass spectra
as was done in Ref.~\cite{LHCb-PAPER-2019-025} is 
complicated by the~fact that 
the~$\Sigmares^{(*)\pm}_{\bquark}\Ppi^{\mp}$ 
and $\Sigmares^{(*)\mp}_{\bquark}\Ppi^{\pm}$ kinematic 
regions overlap in the~range of \lambdabpipi masses used for the~\lbstarstar~fit. 
Separating the~contributions of the~resonant and nonresonant \lbstarstar decays would 
require a~full multidimensional fit in the $\lambdabpipi$, 
$\Lb\pip$ and $\Lb\pim$~masses, which is beyond the~scope of this paper.

\begin{figure}[t]
  \setlength{\unitlength}{1mm}
  \centering
  \begin{picture}(150,150)
    \put(  0, 0){ 
      \includegraphics*[height=150mm,width=150mm,%
      ]{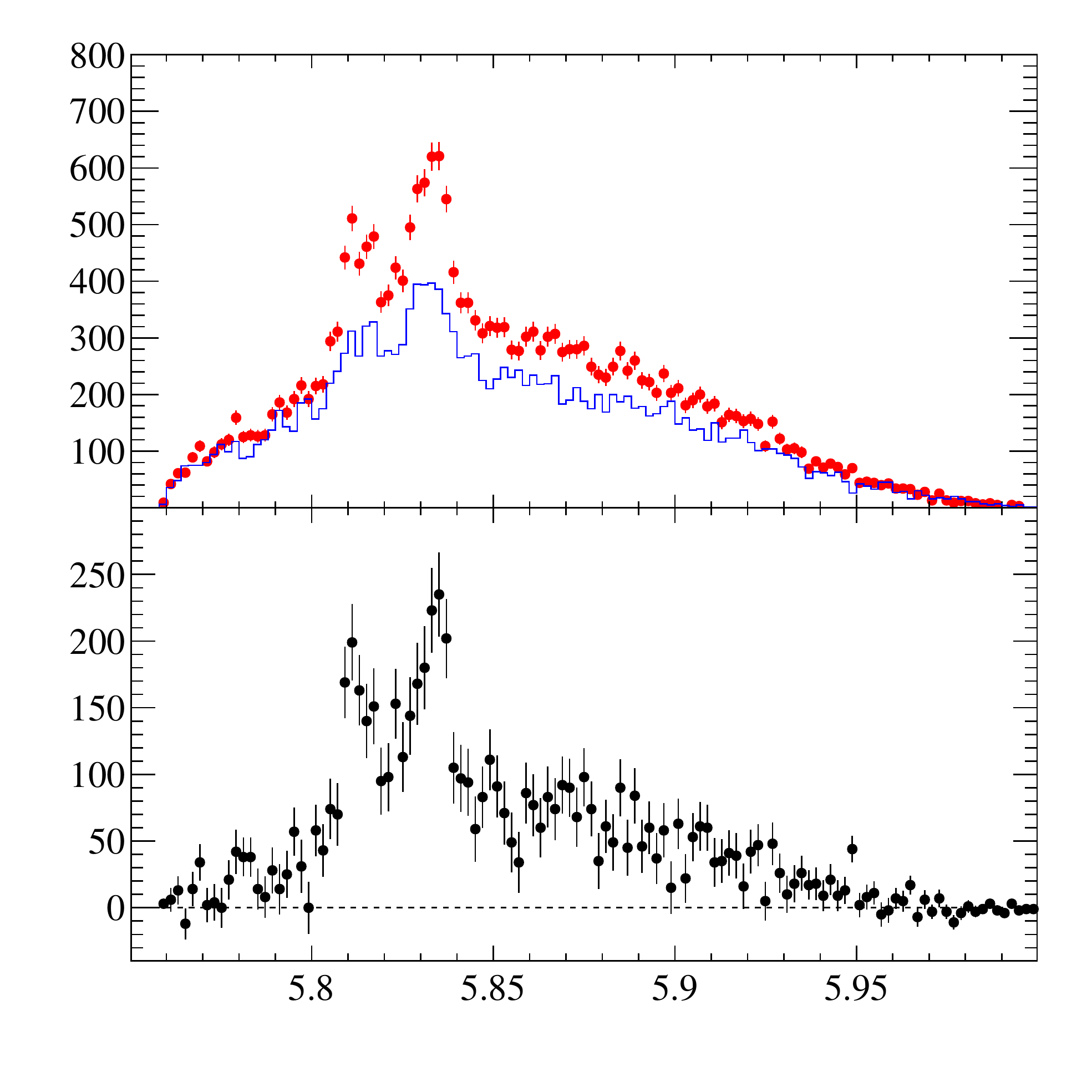}
    }  
    \put(  3,60){\large\begin{sideways}Candidates/$\left(2\mev\right)$\end{sideways}}
     \put(75, 3){\large{$m_{\Lb\Ppi^{\pm}}$}}\put(131, 3) {\large{$\left[\!\gev\right]$}}
    \put(25,135){\large{$\decay{\Lb}{\Lc\pim}$}}
    \put(70,129){{$6.00<m_{\Lb\Ppi\Ppi}<6.14\gev$}}
    \put(117,121.5){\color{red}{\circle*{1.5}}}
    \put(117,119){\color{red}{\line(0,1){5}}}
    \put(114.5,116.6){\color{blue}{\line(1,0){5}}},
    \put(120,120){{$\Lb\pip\pim$}}
    \put(120,115){{$\Lb\Ppi^{\pm}\Ppi^{\pm}$}}
    \put(125,135){\large\lhcb}
  \end{picture}
  \caption{ \small
    (Top)~Spectra of $\Lb\Ppi^{\pm}$~mass with $\decay{\Lb}{\Lc\pim}$
    for $\Lb\pip\pim$~combinations\,(red points with error bars)
    and $\Lb\Ppi^{\pm}\Ppi^{\pm}$~combinations\,(open blue histogram).
    (Bottom)~Difference between $\Lb\Ppi$~mass spectra from 
    $\Lb\pip\pim$ and $\Lb\Ppi^{\pm}\Ppi^{\pm}$~combinations.
    The~structures near $5.81$ and $5.83\gev$
    correspond to
    the~\mbox{$\decay{\Sigma_{\bquark}^{\pm}}{\Lb\Ppi^{\pm}}$} and
    \mbox{$\decay{\Sigma_{\bquark}^{\ast\pm}}{\Lb\Ppi^{\pm}}$}~signals,
    respectively.
  }
  \label{fig:sigmab_spectra}
\end{figure}

The~$\Lb\Ppi^{\pm}$~mass spectra from $\Lb\pip\pim$ and $\Lb\Ppi^{\pm}\Ppi^{\pm}$
combinations with~\mbox{$\decay{\Lb}{\Lc\pim}$} 
from the~\lbstarstar~signal\nobreakdash-enhanced 
region $6.00<m_{\Lb\Ppi\Ppi}<6.14\gev$ are shown 
in Fig.~\ref{fig:sigmab_spectra}. 
The~$\Lb\Ppi^{\pm}$~mass spectrum 
from the~signal \lbstarstar~decays 
is obtained assuming that 
the~$\Lb\Ppi^{\pm}$~spectra 
from the~same\nobreakdash-sign $\Lb\Ppi^{\pm}\Ppi^{\pm}$~combinations 
represent 
the~background. 
The~background-subtracted spectrum is consistent with the
presence of relatively small contributions 
from  \mbox{$\decay{\lbstarstar}{\Sigma_{\bquark}^{\pm}\Ppi^{\mp}}$}
and 
\mbox{$\decay{\lbstarstar}{\Sigma_{\bquark}^{\ast\pm}\Ppi^{\mp}}$} decays
and a~dominant contribution 
from nonresonant
\mbox{$\decay{\lbstarstar}{\Lb\pip\pim}$}~decays.

\section{Analysis of the~low\nobreakdash-mass region} 

\begin{figure}[t]
  \setlength{\unitlength}{1mm}
  \centering
  \begin{picture}(150,150)
    \put(  0, 0){ 
      \includegraphics*[height=150mm,width=150mm,%
      ]{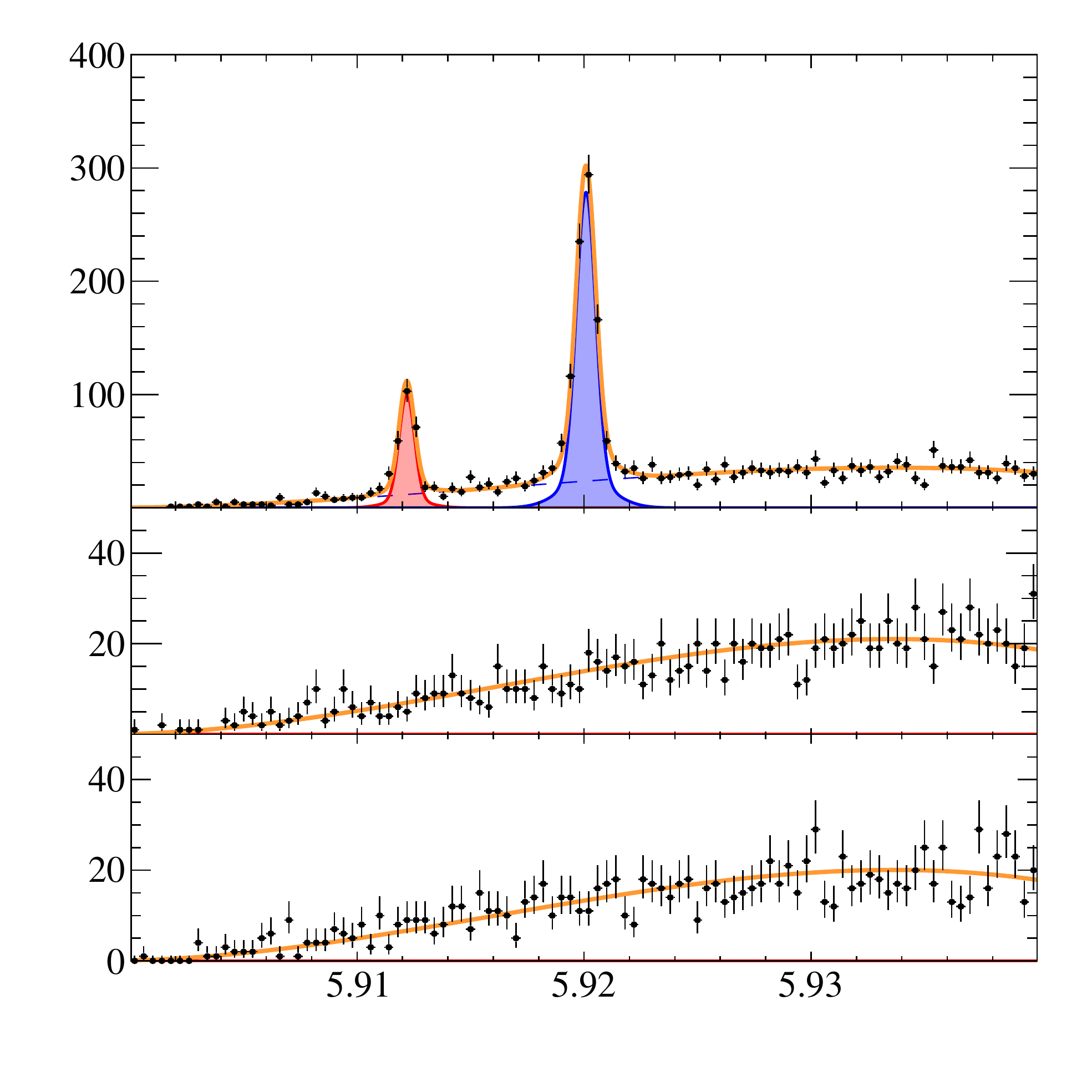}
    }  
    \put(25,130) {\begin{tikzpicture}[x=1mm,y=1mm]\filldraw[fill=red!35!white,draw=red,thick]  (0,0) rectangle (10,1.5);\end{tikzpicture} }
    \put(25,126) {\begin{tikzpicture}[x=1mm,y=1mm]\filldraw[fill=blue!35!white,draw=blue,thick]  (0,0) rectangle (10,1.5);\end{tikzpicture} }
    \put(25,122) {\color[rgb]{0.00,0.00,1.00} {\hdashrule[0.5ex][x]{10mm}{0.3mm}{1.5mm 1.0mm} } } 
    \put(25,118) {\color[rgb]{1.00,0.65,0.00} {\hdashrule[0.5ex][x]{10mm}{0.8mm}{1.0mm 0.0mm} } } 
    \put(38,130) {\small{$\Lambda_{\bquark}\mathrm{(5912)}^0$}}
    \put(38,126) {\small{$\Lambda_{\bquark}\mathrm{(5920)}^0$}}
    \put(38,122) {\small{background}}
    \put(38,118) {\small{total}}
    \put(5 ,73){\large\begin{sideways}Candidates/$\left(0.4\mev\right)$\end{sideways}}
    \put(75, 3){\large{$m_{\Lb\Ppi\Ppi}$}}\put(131, 3) {\large{$\left[\!\gev\right]$}}
    \put(55,135){\large{$\decay{\Lb}{\Lc\pim}$}}
    \put(25,135){\large{$\Lb\pip\pim$}}
    \put(25, 73){\large{$\Lb\pip\pip$}}
    \put(25, 42){\large{$\Lb\pim\pim$}}
    \put(125,135){\large\lhcb}
  \end{picture}
  \caption { \small
     Mass spectra of selected 
    (top)~$\Lb\pip\pim$,
    (middle)~$\Lb\pip\pip$ and 
    (bottom)~$\Lb\pim\pim$~combinations for 
   for the~$\decay{\Lb}{\Lc\pim}$~sample.
    A~simultaneous fit, described in the~text, is superimposed.
  }
  \label{fig:fit_lowmass_lcpi}
\end{figure}

\begin{figure}[t]
  \setlength{\unitlength}{1mm}
  \centering
  \begin{picture}(150,150)
    \put(  0, 0){ 
      \includegraphics*[height=150mm,width=150mm,%
      ]{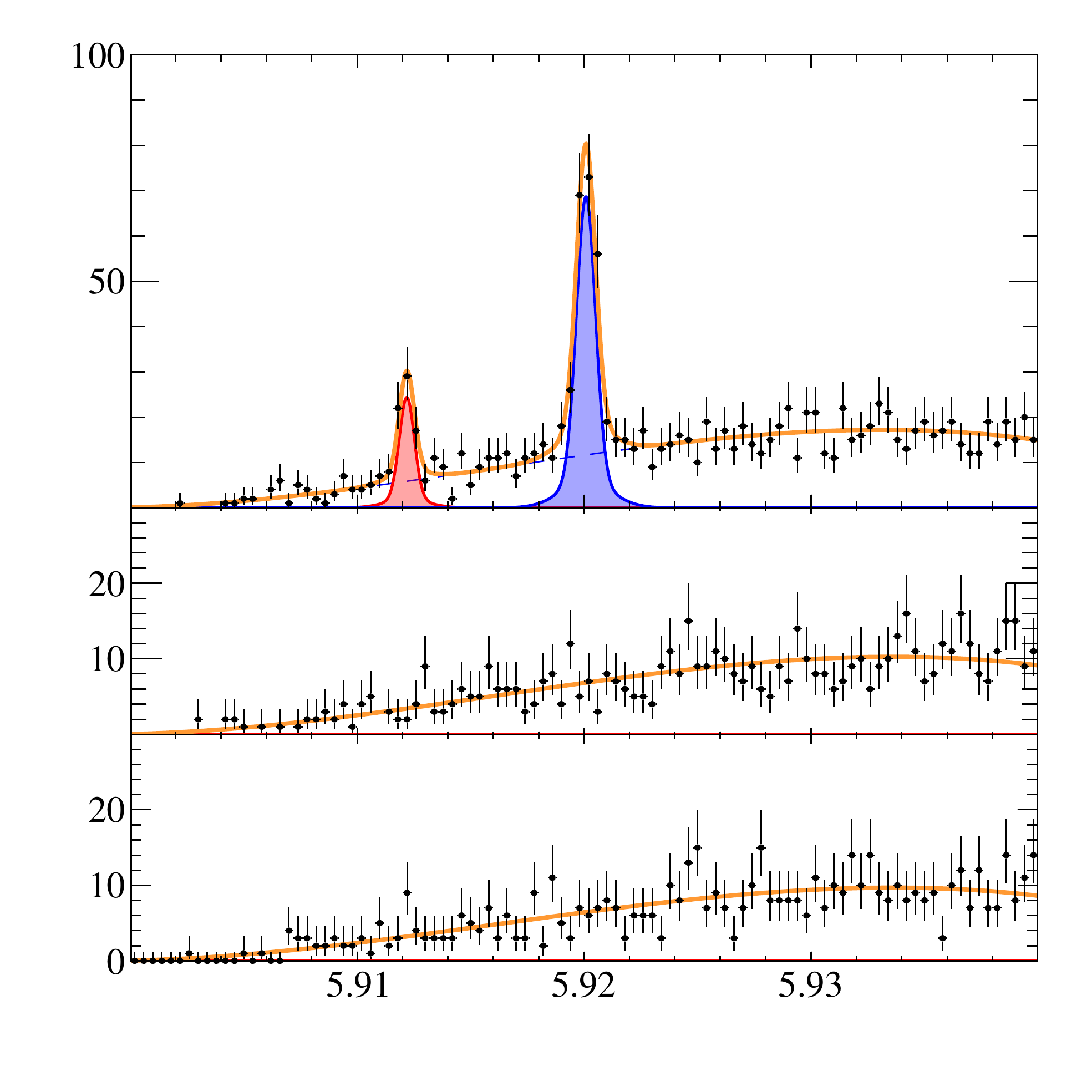}
    }  
    \put(25,130) {\begin{tikzpicture}[x=1mm,y=1mm]\filldraw[fill=red!35!white,draw=red,thick]  (0,0) rectangle (10,1.5);\end{tikzpicture} }
    \put(25,126) {\begin{tikzpicture}[x=1mm,y=1mm]\filldraw[fill=blue!35!white,draw=blue,thick]  (0,0) rectangle (10,1.5);\end{tikzpicture} }
    \put(25,122) {\color[rgb]{0.00,0.00,1.00} {\hdashrule[0.5ex][x]{10mm}{0.3mm}{1.5mm 1.0mm} } } 
    \put(25,118) {\color[rgb]{1.00,0.65,0.00} {\hdashrule[0.5ex][x]{10mm}{0.8mm}{1.0mm 0.0mm} } } 
    \put(38,130) {\small{$\Lambda_{\bquark}\mathrm{(5912)}^0$}}
    \put(38,126) {\small{$\Lambda_{\bquark}\mathrm{(5920)}^0$}}
    \put(38,122) {\small{background}}
    \put(38,118) {\small{total}}
    \put(5 ,73){\large\begin{sideways}Candidates/$\left(0.4\mev\right)$\end{sideways}}
    \put(75, 3){\large{$m_{\Lb\Ppi\Ppi}$}}\put(131, 3) {\large{$\left[\!\gev\right]$}}
    \put(55,135){\large{$\decay{\Lb}{\jpsi\proton\Km}$}}
    \put(25,135){\large{$\Lb\pip\pim$}}
    \put(25, 73){\large{$\Lb\pip\pip$}}
    \put(25, 42){\large{$\Lb\pim\pim$}}
    \put(125,135){\large\lhcb}
  \end{picture}
  \caption { \small
    Mass spectra of selected 
    (top)~$\Lb\pip\pim$,
    (middle)~$\Lb\pip\pip$ and 
    (bottom)~$\Lb\pim\pim$~combinations for 
   for the~$\decay{\Lb}{\jpsi\proton\Km}$~sample.
    A~simultaneous fit, described in the~text, is superimposed.
   }
  \label{fig:fit_lowmass_psipk}
\end{figure}

The~$\Lb\Ppi\Ppi$~mass spectra  in the~low\nobreakdash-mass region $m_{\Lb\Ppi\Ppi}<5.94\gev$ 
for \lblcpi and \lbjpsipk samples are shown in Figs.~\ref{fig:fit_lowmass_lcpi} 
and \ref{fig:fit_lowmass_psipk}, respectively. 
These~distributions are used to measure the~properties of 
the~\lboneplow and \lbonephigh states. 
A~simultaneous binned fit, with narrow bins of 50\kev width,
is performed to 
the~six distributions with the~sum of 
the~two resonance 
components\,(in \lambdabpipi combinations only)
and  the~combinatorial background 
component\,(in all six distributions). 
The combinatorial component is parameterised with a~product of 
the~three\nobreakdash-body phase\nobreakdash-space  
function and a~positive polynomial function. 
The~resonant components are given by relativistic $S$-wave Breit--Wigner lineshapes convolved with 
the~resolution function obtained from simulation. 
The~shape of the~combinatorial background  
is assumed to be the~same in the~opposite\nobreakdash-sign \lambdabpipi
and same\nobreakdash-sign \lambdabpipiSS spectra, but is allowed to differ for the~\lblcpi and \lbjpsipk samples.
The results of the combined fit are presented in Table~\ref{tab:lowmass_results}. 
The~natural widths of the~\lboneplow
and \lbonephigh states are consistent with zero. 

\begin{table}[b]
  \centering
  \caption{\small
    Results of the combined fit to the low-mass $\Lb\Ppi\Ppi$ spectra. 
  } \label{tab:lowmass_results}
  \vspace*{3mm}
  \begin{tabular*}{0.85\textwidth}{@{\hspace{5mm}}l@{\extracolsep{\fill}}lcc@{\hspace{5mm}}}
    &   & $\decay{\Lb}{\Lc\pim}$ & $\decay{\Lb}{\jpsi\proton\Km}$
    \\[1mm]
    \hline 
    \\[-2mm]
    $N_{\Lambda_{\bquark}\mathrm{(5912)}^0}$ & & $234\pm17$                     
                                           & $\phantom{0}57\pm9\phantom{0}$\\
    $N_{\Lambda_{\bquark}\mathrm{(5920)}^0}$ & & $843\pm33$ 
                                           & $204\pm17$ \\
    $\Delta m_{\Lambda_{\bquark}\mathrm{(5912)}^0}$ & $\left[\!\mev\right]$ & \multicolumn{2}{c}{$\phantom{0}292.582 \pm 0.029$} \\ 
    $\Delta m_{\Lambda_{\bquark}\mathrm{(5920)}^0}$ & $\left[\!\mev\right]$ & \multicolumn{2}{c}{$\phantom{0}300.479 \pm 0.019$} \\ 
    $m_{\Lambda_{\bquark}\mathrm{(5920)}^0} - m_{\Lambda_{\bquark}\mathrm{(5912)}^0}$
    & $\left[\!\mev\right]$ & \multicolumn{2}{c}{$\phantom{000}7.896\pm0.034$}
  \end{tabular*}
\end{table}

\section{Systematic uncertainties}

The systematic uncertainties of
the~mass and the~width of 
the \lbstarstar state and of the~masses of the~\lboneplow and \lbonephigh states 
are summarised in Table~\ref{tab:syst_total}.

\begin{table}[t]
  \centering
  \caption{ \small
    Summary of systematic uncertainties for 
    the~mass difference  with respect to the~ground state~\Lb
    and natural width
    of the~\lbstarstar~state and the~mass\nobreakdash-differences 
    for 
    the~$\Lambda_{\bquark}\mathrm{(5912)}^0$ and 
     $\Lambda_{\bquark}\mathrm{(5920)}^0$~states, 
     $\Delta m_{\Lambda_{\bquark}\mathrm{(1P)}^0}$.
  } \label{tab:syst_total}
  \vspace*{1mm}
  \begin{tabular*}{0.75\textwidth}{@{\hspace{5mm}}l@{\extracolsep{\fill}}ccc@{\hspace{5mm}}}
    \multirow{2}{*}{Source} 
    &  $\Delta m_{\lbstarstar}$
    &  $\Gamma_{\lbstarstar}$
    &  $\Delta m_{\Lambda_{\bquark}\mathrm{(1P)}^0}$.
    \\
    & $\left[\!\mev\right]$
    & $\left[\!\mev\right]$
    & $\left[\!\mev\right]$
    \\[1mm]
    \hline 
    \\[-4mm]
    Fit model                                        &       &       & \\ 
    ~~Signal parameterisation                        & 0.50  & 1.50  & \\ 
    ~~Background parameterisation                    & 0.03  & 0.25  & \\ 
    ~~Fit range                                      & 0.10  & 0.30  & \\ 
    ~~$\Lambda_{\bquark}\mathrm{(1D)}^0$~parameters  &       &       & \\ 
    Momentum scale uncertainty                       & 0.08  & --    & 0.010  
    \\[1mm]
    \hline 
    \\[-4mm]
    Sum in quadrature    & 0.52 & 1.55  &  0.010      
  \end{tabular*}
  \vspace*{3mm}
\end{table}

A large uncertainty in the~measurement of the~\lbstarstar parameters  comes from the~parameterisation of the~\lbstarstar signal distribution. The fit function from Eq.~\eqref{eq:signal} describes three\nobreakdash-body phase\nobreakdash-space decays, while Fig.~\ref{fig:sigmab_spectra} 
suggests some contribution from decays via 
the~intermediate $\Sigma_{\bquark}^{(\ast)\pm}\Ppi^{\mp}$~states.
To~assess the associated systematic uncertainty, 
the~fit is repeated using a~more complicated 
function that in addition to 
nonresonant decays,  
accounts for the~P\nobreakdash-wave decays via 
an~intermediate $\Sigma_{\bquark}^{(\ast)\pm}\Ppi^{\mp}$~state, 
but ignores interference effects,
    constructed using
    the~three\nobreakdash-particle unitarity
    constraint approximated in
    the~quasi\nobreakdash-two\nobreakdash-body
    interaction model~\cite{Mikhasenko:2019vhk}
\begin{equation}
  \mathfrak{S}^{\prime}\left(m|m_0,\Gamma_{\mathrm{NR}},
  \Gamma_{\Sigma_{\bquark}\Ppi} , \Gamma_{\Sigma_{\bquark}^{\ast}\Ppi}\right) \propto 
    \dfrac{  \Gamma\left(m\right)}
    { \left( m^2_0 - m^2 \right)^2 + m_0^2 \Gamma^2\left(m\right)} \, , 
\end{equation}
where the~mass-dependent width $\Gamma\left(m\right)$ is defined as 
\begin{equation*}
  \Gamma \left( m\right)  = 
    \Gamma_{\mathrm{NR}} \dfrac{\Prho_{3}(m)}{ \Prho_3(m_0)} + 
    \Gamma_{\Sigma_{\bquark}\Ppi} \dfrac{ \Prho_{\Sigma_{\bquark}\Ppi}(m)}{\Prho_{\Sigma_{\bquark}\Ppi}(m_0) } + 
    \Gamma_{\Sigma_{\bquark}^{\ast}\Ppi}\dfrac{ \Prho_{\Sigma_{\bquark}^{\ast}\Ppi}(m)}{  \Prho_{\Sigma_{\bquark}^{\ast}\Ppi}(m_0)}    \,. 
\end{equation*}
The~quasi-two-body phase-space functions $\Prho_{\Sigma_{\bquark}^{(*)}\Ppi}(m)$ for the~decays via the~intermediate 
$\Sigma_{\bquark}\Ppi$
and 
$\Sigma_{\bquark}^{\ast}\Ppi$~states are 
\begin{eqnarray}
     \Prho_{\Sigma_{\bquark}^{(*)}\Ppi}\left(m\right) & = &
  \int\limits_{(m_{\Ppi}+m_{\Lb})^2}^{(m-m_{\Ppi})^2}
  \dfrac{ 
  \left(
      \dfrac{2p}{m}\,
      \dfrac{2q}{\sqrt{s}}\,
      \dfrac{R^2p^2}{1+R^2p^2}\,
      \dfrac{R^2q^2}{1+R^2q^2}\right)
  }{ ( m^2_{\Sigma_{\bquark}^{(\ast)}} - s)^2 + 
    m^2_{\Sigma_{\bquark}^{(\ast)}} 
    \Gamma^{\prime2}_{\Sigma_{\bquark}^{(\ast)}} \left( s \right)  }
  ds\,, \nonumber \\
  \Gamma^{\prime}_{\Sigma_{\bquark}^{(\ast)}} \left( s  \right)   
  & = & 
  \Gamma_{\Sigma_{\bquark}^{(\ast)}} \dfrac{m_{\Sigma_{\bquark}^{(\ast)}}}{\sqrt{s}}
   \left( \dfrac{q}{ q_0 } \right)^3 
   \left(  \dfrac{ 1+R^2 q^2} { 1 + R^2q^2_0} \right)^2\,, \nonumber
\end{eqnarray}
where $s$ stands for a squared mass of the~$\Lb\Ppi$~pair forming 
the~$\Sigma_{\bquark}^{(\ast)}$~resonance, 
$p$~denotes the~momenta of the~pion in the~P\nobreakdash-wave decay 
$\decay{\lbstarstar}{\Sigma_{\bquark}^{(\ast)}\Ppi}$,
$q$~denotes the~momenta of the~pion in the~decay
$\decay{\Sigma_{\bquark}^{(*)}}{\Lb\Ppi}$,
$q_0$ is the value of $q$ at $s=m_{\Sigma_{\bquark}^{(*)}}$,
$R=3.5\gev^{-1}$ corresponds to the~breakup momentum of
the~P\nobreakdash-wave Blatt--Weisskopf centrifugal 
barrier factor~\cite{Blatt:1952ije}, 
$m_{\Sigma_{\bquark}^{(\ast)}}$~and 
$\Gamma_{\Sigma_{\bquark}^{(\ast)}}$~are known
mass and width  of 
the~$\Sigma_{\bquark}^{(\ast)}$~states~\cite{LHCb-PAPER-2018-032}.
The~function is reparameterised as 
\begin{eqnarray*}
  \Gamma_{\mathrm{NR}}              & = & \left( 1 - \upalpha - \upbeta \right)\Gamma\,,\\
  \Gamma_{\Sigma_{\bquark}\Ppi       } & = & \upalpha\,\Gamma \,,\\
  \Gamma_{\Sigma_{\bquark}^{\ast}\Ppi} & = & \upbeta\,\Gamma \,,
\end{eqnarray*}
where the non\nobreakdash-negative parameters 
$\upalpha$ and $\upbeta$ account for  
the~relative contributions 
from the~
\mbox{$\decay{\lbstarstar}{\Sigma_{\bquark}^{\pm}\Ppi^{\mp}}$} and 
\mbox{$\decay{\lbstarstar}{\Sigma_{\bquark}^{\ast\pm}\Ppi^{\mp}}$}~decays, 
respectively.
A~series of fits is performed with 
parameters $\upalpha$ and $\upbeta$ varied within 
the~ranges $0 \le \upalpha<0.2$, $0 \le \upbeta<0.2$,
and $\upalpha+\upbeta\le0.3$, 
consistent with Fig.~\ref{fig:sigmab_spectra}.
The~mass of the~$\lbstarstar$~state is 
found to be very stable with respect to such 
variations. The~fitted mass does
not change more than 0.5\mev while  
the~fitted width increases up to~1.5\mev. 
These values are taken as systematic 
uncertainties due to the~signal parameterisation.
The~nominal fit does not take 
the~variations of 
the~detector efficiency with 
the~$\lambdabpipi$ mass into account. 
An~alternative fit is performed where the~signal shape is multiplied by 
the~efficiency function obtained from simulation. 
The~difference with the nominal fit is added 
to the~uncertainty on the~signal parameterisation. 
Alternative parameterisations of the~detector resolution functions, 
namely 
a~symmetric variant of an~Apollonios function~\cite{Santos:2013gra},
a~double\nobreakdash-sided Crystal Ball function~\cite{Skwarnicki:1986xj}, 
a~modified Novosibirsk function~\cite{bukin2007,Lees:2011gw},
a~Student's $t$\nobreakdash-distribution
and 
a~hyperbolic secant function, cause negligible variation for the~measured 
mass and width of the~\lbstarstar~state.
The~signal parameterisation uncertainty in the~measurement of the~masses of the~low\nobreakdash-mass states is negligible. 

The~uncertainty in the~combinatorial background shape parameterisation is accounted for by varying 
the~degree of the~polynomial functions from 3 to 4. 
The~uncertainty in the~$\Sigmab\Ppi$ and $\Sigmabstar\Ppi$ background functions 
is evaluated by modifying the~parameters of the~exponential parameterisation within 
the~limits allowed by the~fits to the~background\nobreakdash-subtracted $\Sigmares_{\bquark}^{(*)}\Ppi$ spectra. 
In order to~assess a possible sensitivity of the~fit parameters to the~features of the~background 
shape not accounted for by the~variations mentioned above,
fits are performed in narrower and broader $\Lb\Ppi\Ppi$ regions and variations are included as an~additional source of systematic uncertainty. 

To assess the effect of the~fixed parameters of 
the~narrow \lbonedlow
and \lbonedhigh states from 
the~previous analysis~\cite{LHCb-PAPER-2019-025}
in the higher-mass fit, 
the~fits are performed with the~masses and the~widths of 
each of the~two states 
left free to vary one by~one. 
The~resulting variations of the~\lbstarstar parameters are found to be negligible. 

The effect of the calibration of the momentum scale is 
evaluated by varying the scale within its known 
uncertainty~\cite{LHCb-PAPER-2012-012,LHCb-PAPER-2019-025,LHCb-PAPER-2013-011}. 
All~systematic uncertainties for the~mass difference 
\mbox{$m_{\Lambda_{\bquark}\mathrm{(5920)}^0} - m_{\Lambda_{\bquark}\mathrm{(5912)}^0}$} 
are found to be negligible.

The~upper limits on the~natural widths of the~\lboneplow and \lbonephigh
states are obtained by performing profile likelihood scans. 
In~the calculation of the likelihood, 
the~uncertainties in the~knowledge of mass resolution are included by using various 
resolution models, as listed above, and by varying the~mass\nobreakdash-resolution scaling factor obtained from 
simulations within $5\%$~\cite{LHCb-PAPER-2017-036,LHCb-PAPER-2019-005,LHCb-PAPER-2019-025} 
and the~maximum upper limits across 
all variations are reported.

\section{Results and summary}

Using the~LHCb data set taken in 2011--2018, 
corresponding to an~integrated luminosity of 
9\invfb collected in $\proton\proton$~collisions at 
centre-of-mass energies of 7, 8 and 13\tev, 
the~$\Lb\pip\pim$~mass spectrum 
is studied with $\Lb$~baryons reconstructed in the~\mbox{$\decay{\Lb}{\Lc\pim}$}
and~\mbox{$\decay{\Lb}{\jpsi\proton\Km}$}~decay modes. 
A~new broad resonance-like state is observed with 
a~statistical significance exceeding  14 and 7~standard 
deviations for $\Lb\pip\pim$~samples reconstructed 
using the~\mbox{$\decay{\Lb}{\Lc\pim}$}
and~\mbox{$\decay{\Lb}{\jpsi\proton\Km}$}~decay modes, respectively.
The~mass difference with respect to the~\Lb~mass 
and natural width of the state are determined from 
a~combined fit to both samples and are found to~be 
\begin{eqnarray*}
  \Delta m_{\lbstarstar}      & = &   452.7 \pm 2.9 \pm 0.5 \mev\,, \\
  \Gamma_{\lbstarstar} & = &  \phantom{0}72\phantom{.0}\pm 11\phantom{.} \pm 2\phantom{.0} \mev\,,
\end{eqnarray*}
where the first uncertainty is statistical and the second systematic.
Taking the~mass of the~\Lb~baryon 
\mbox{$m_{\Lb}=5619.62 \pm 0.16 \pm 0.13\mev$}~\cite{LHCb-PAPER-2017-011}, 
obtained by a~combination of measurements at  the~LHCb experiment 
in 
$\decay{\Lb}{\Pchi_{\cquark1,2}\proton\Km}$~\cite{LHCb-PAPER-2017-011}, 
$\decay{\Lb}{\Ppsi\mathrm{(2S)}\proton\Km}$,
$\decay{\Lb}{\jpsi\pip\pim\proton\Km}$~\cite{LHCb-PAPER-2015-060}
and 
$\decay{\Lb}{\jpsi\Lambda}$~decay modes~\cite{LHCb-PAPER-2011-035,LHCb-PAPER-2012-048}, 
and accounting for the~correlated systematic uncertainty,  
the~mass of the~\lbstarstar state is found to~be 
\begin{equation*}
  m_{\lbstarstar}  =  6072.3 \pm 2.9 \pm 0.6 \pm 0.2 \mev\,,
\end{equation*}
where the last uncertainty is due to that on  the~mass of the~\Lb~baryon. 
The~new resonance  is consistent with the~broad excess of events  
reported by the~CMS collaboration~\cite{Sirunyan:2020gtz} 
and the~measured mass and width agree with expectations  
for the~$\Lambda_{\bquark}\mathrm{(2S)}^{0}$~state~\cite{Capstick:1986bm,Roberts:2007ni,Ebert:2011kk,Yamaguchi:2014era,Chen:2018vuc}.
  
  Several excited $\Sigma_{\bquark}\mathrm{(1P)}$~states are expected with 
  a~mass close to the~measured value, but 
  the~partial decay widths for $\Sigma_{\bquark}\mathrm{(1P)}$~states 
  into  $\Lb\Ppi\Ppi$~are predicted  to be very small~\cite{Mu:2014iaa}.
If~the~observed broad peak corresponds to 
the  $\Sigma_{\bquark}\mathrm{(1P)}^{(\ast)0}$~state,
two~peaks with similar masses and widths and significantly 
larger yields should be visible in the~$\Lb\Ppi^{\pm}$~mass
spectra due to decays of the~charged isospin partners \mbox{$\decay{\Sigma_{\bquark}\mathrm{(1P)}^{(\ast)\pm}}{\Lb\Ppi^{\pm}}$}.
However, no signs of states with such a~mass and width, 
and large production yields are observed in the~analysis of the~$\Lb\Ppi^{\pm}$~mass spectra;
the~observed $\Sigma_{\bquark}\mathrm{(6097)}^{\pm}$~states
have significantly smaller natural width 
and relatively small yields~\cite{LHCb-PAPER-2018-032}.  
It~cannot be excluded that the~observed broad structure 
corresponds to a~superposition of  more than one narrow states, but 
the~interpretation of these states as excited 
$\Sigma_{\bquark}$~resonances is disfavoured.  
The~mass differences for 
the~$\Lambda_{\bquark}\mathrm{(5912)}^0$ and 
$\Lambda_{\bquark}\mathrm{(5920)}^0$~states with respect to 
the~mass of the~\Lb~baryon are measured to be 
\begin{eqnarray*}
  \Delta m_{\Lambda_{\bquark}\mathrm{(5912)}^0} & = & 292.589 \pm 0.029 \pm 0.010 \mev\,, \\
  \Delta m_{\Lambda_{\bquark}\mathrm{(5920)}^0} & = & 300.492 \pm 0.019 \pm 0.010 \mev\,, 
\end{eqnarray*}
and the corresponding masses are
\begin{eqnarray*}
  m_{\Lambda_{\bquark}\mathrm{(5912)}^0} & = & 5912.21 \pm 0.03 \pm 0.01 \pm0.21 \mev\,, \\
  m_{\Lambda_{\bquark}\mathrm{(5920)}^0} & = & 5920.11 \pm 0.02 \pm 0.01 \pm 0.21 \mev\,, 
\end{eqnarray*}
where the last uncertainty is due to imprecise knowledge of the~\Lb~mass.
The~mass splitting between the narrow states is 
\begin{equation*}
m_{\Lambda_{\bquark}\mathrm{(5920)}^0} - 
m_{\Lambda_{\bquark}\mathrm{(5912)}^0} = 7.896 \pm 0.034 \mev\,.
\end{equation*}
The following upper limits on the~natural widths are obtained: 
\begin{eqnarray*}
  \Gamma_{\Lambda_{\bquark}\mathrm{(5912)}^0} & < & 0.25\,(0.28) \mev\,, \\
  \Gamma_{\Lambda_{\bquark}\mathrm{(5920)}^0} & < & 0.19\,(0.20) \mev\,, 
\end{eqnarray*}
at 90\%\,(95\%) confidence level, respectively. 
The~measurements of the~parameters of the~\lboneplow and \lbonephigh
states are about four times more precise and supersede those reported in Ref.~\cite{LHCb-PAPER-2012-012}.

\section*{Acknowledgements}
%
%
\noindent We express our gratitude to our colleagues in the~CERN
accelerator departments for the excellent performance of the LHC. We
thank the technical and administrative staff at the LHCb
institutes.
We acknowledge support from CERN and from the national agencies:
CAPES, CNPq, FAPERJ and FINEP\,(Brazil); 
MOST and NSFC\,(China); 
CNRS/IN2P3\,(France); 
BMBF, DFG and MPG\,(Germany); 
INFN\,(Italy); 
NWO\,(Netherlands); 
MNiSW and NCN\,(Poland); 
MEN/IFA\,(Romania); 
MSHE\,(Russia); 
MinECo\,(Spain); 
SNSF and SER\,(Switzerland); 
NASU\,(Ukraine); 
STFC\,(United Kingdom); 
DOE NP and NSF\,(USA).
We~acknowledge the computing resources that are provided by CERN, 
IN2P3\,(France), 
KIT and DESY\,(Germany), 
INFN\,(Italy), 
SURF\,(Netherlands),
PIC\,(Spain), 
GridPP\,(United Kingdom), 
RRCKI and Yandex LLC\,(Russia), 
CSCS\,(Switzerland), 
IFIN\nobreakdash-HH\,(Romania), 
CBPF\,(Brazil),
PL\nobreakdash-GRID\,(Poland) and OSC\,(USA).
We~are indebted to the communities behind the~multiple open\nobreakdash-source
software packages on which we depend.
Individual groups or members have received support from
AvH Foundation\,(Germany);
EPLANET, Marie Sk\l{}odowska\nobreakdash-Curie Actions and ERC\,(European Union);
ANR, Labex P2IO and OCEVU, and R\'{e}gion Auvergne\nobreakdash-Rh\^{o}ne\nobreakdash-Alpes\,(France);
Key Research Program of Frontier Sciences of CAS, CAS PIFI, and the Thousand Talents Program\,(China);
RFBR, RSF and Yandex LLC\,(Russia);
GVA, XuntaGal and GENCAT\,(Spain);
the Royal Society
and the~Leverhulme Trust\,(United Kingdom).





\clearpage
\addcontentsline{toc}{section}{References}
\setboolean{inbibliography}{true}
\bibliographystyle{LHCb}
\bibliography{standard,local,LHCb-PAPER,LHCb-CONF,LHCb-DP,LHCb-TDR}

\newpage

\centerline
{\large\bf LHCb collaboration}
\begin
{flushleft}
\small
R.~Aaij$^{31}$,
C.~Abell{\'a}n~Beteta$^{49}$,
T.~Ackernley$^{59}$,
B.~Adeva$^{45}$,
M.~Adinolfi$^{53}$,
H.~Afsharnia$^{9}$,
C.A.~Aidala$^{80}$,
S.~Aiola$^{25}$,
Z.~Ajaltouni$^{9}$,
S.~Akar$^{66}$
% ...

{\footnotesize \it

% ...
$ ^{9}$Universit{\'e} Clermont Auvergne, CNRS/IN2P3, LPC, Clermont-Ferrand, France\\
% ...
$ ^{25}$INFN Sezione di Milano, Milano, Italy\\
% ...
$ ^{31}$Nikhef National Institute for Subatomic Physics, Amsterdam, Netherlands\\
% ...
$ ^{45}$Instituto Galego de F{\'\i}sica de Altas Enerx{\'\i}as (IGFAE), Universidade de Santiago de Compostela, Santiago de Compostela, Spain\\
% ...
$ ^{49}$Physik-Institut, Universit{\"a}t Z{\"u}rich, Z{\"u}rich, Switzerland\\
% ...
$ ^{53}$H.H. Wills Physics Laboratory, University of Bristol, Bristol, United Kingdom\\
% ...
$ ^{59}$Oliver Lodge Laboratory, University of Liverpool, Liverpool, United Kingdom\\
% ...
$ ^{66}$Los Alamos National Laboratory (LANL), Los Alamos, United States\\
$ ^{67}$Syracuse University, Syracuse, NY, United States\\
% ...
$ ^{80}$University of Michigan, Ann Arbor, United States, associated to $^{67}$\\
\bigskip
% ...
\medskip
}
\end{flushleft}

\end{document}